\documentclass[aps,prl,twocolumn,superscriptaddress,a4paper,10pt]{revtex4-1}
\usepackage{graphicx}
\usepackage{dcolumn}
\usepackage{bm}
\usepackage{amsmath}
 
\graphicspath{{/}}
\usepackage{color}
\usepackage{cancel}
\usepackage{BOONDOX-cal} 

\usepackage{upgreek}
\usepackage[latin1]{inputenc}

\begin{document}

\title{Resonant magnetic induction tomography of a magnetized sphere}

\author{A. Gloppe}
\email[]{arnaud.gloppe@qc.rcast.u-tokyo.ac.jp}
\affiliation{Research Center for Advanced Science and Technology (RCAST), The University of Tokyo, Meguro-ku, Tokyo 153-8904, Japan}
\author{R. Hisatomi}
\affiliation{Research Center for Advanced Science and Technology (RCAST), The University of Tokyo, Meguro-ku, Tokyo 153-8904, Japan}
\author{Y. Nakata}
\affiliation{Research Center for Advanced Science and Technology (RCAST), The University of Tokyo, Meguro-ku, Tokyo 153-8904, Japan}
\author{Y. Nakamura}
\affiliation{Research Center for Advanced Science and Technology (RCAST), The University of Tokyo, Meguro-ku, Tokyo 153-8904, Japan}
\affiliation{Center for Emergent Matter Science (CEMS), RIKEN, Wako, Saitama 351-0198, Japan}
\author{K. Usami}
\affiliation{Research Center for Advanced Science and Technology (RCAST), The University of Tokyo, Meguro-ku, Tokyo 153-8904, Japan}

\date{\today}

\begin{abstract}
We demonstrate the structural imaging of magnetostatic spin-wave modes hosted in a millimeter-sized ferromagnetic sphere. Unlike for low-dimensional magnetic materials, there is no prior technique to image these modes in bulk magnetized solid of revolution.  Based on resonant magnetic induction tomography in the microwave range, our approach ensures the robust identification of these non-trivial spin-wave modes by establishing their azimuthal and polar dependences, starting point of magnonic fundamental studies and hybrid systems with complex spin textures well beyond the uniform precession mode.
\end{abstract}
\maketitle

\paragraph*{}
Macroscopic magnetically ordered structures, such as Yttrium Iron Garnet (YIG) millimetric spheres, are solid supports of extended collective spin excitations, \textit{magnons}~\cite{Gurevich1996, Stancil2009, Kruglyak2010}, that can be cooled down to their quantum ground state and coherently coupled to a superconducting quantum bit through a microwave cavity~\cite{Tabuchi2015}. A coherent optical control of magnons in the quantum regime could enable the efficient transduction of optical and microwave photons~\cite{Hisatomi2016}, opening the way to quantum telecommunications between superconducting quantum computers~\cite{Kimble2008,Wehner2018} as well as quantum-noise limited microwave amplifiers~\cite{Barzanjeh2015}. The study of the interactions of magnons and photons in an optical cavity, or \textit{cavity optomagnonics}, in a solid-state matrix has been initiated by the first observations of magnon-induced Brillouin light scattering involving the uniform precession spin-wave mode and optical whispering gallery modes of a YIG sphere~\cite{Osada2016,Haigh2016,Zhang2016}. Higher-order magnetostatic spin-wave modes, with a variety of orbital angular momenta and spin textures, extend the richness of this hybrid system. In particular, the exchange of orbital angular momentum between magnons and optical photons has been experimentally demonstrated recently~\cite{Osada2018,Haigh2018}. \color{black}The related selection rules allow a controlled non-reciprocal scattering, dependent on the interacting spin wave, potentially leading to the development of a new class of chiral devices~\cite{Lodahl2017}. 
The optomagnonic coupling, faint with the uniform precession mode, will be optimized for high-order modes whose spatial distribution localizes more and more towards the resonator boundaries where the optical whispering gallery modes spread~\cite{Sharma2017}.

\begin{figure}
\includegraphics[scale=0.95]{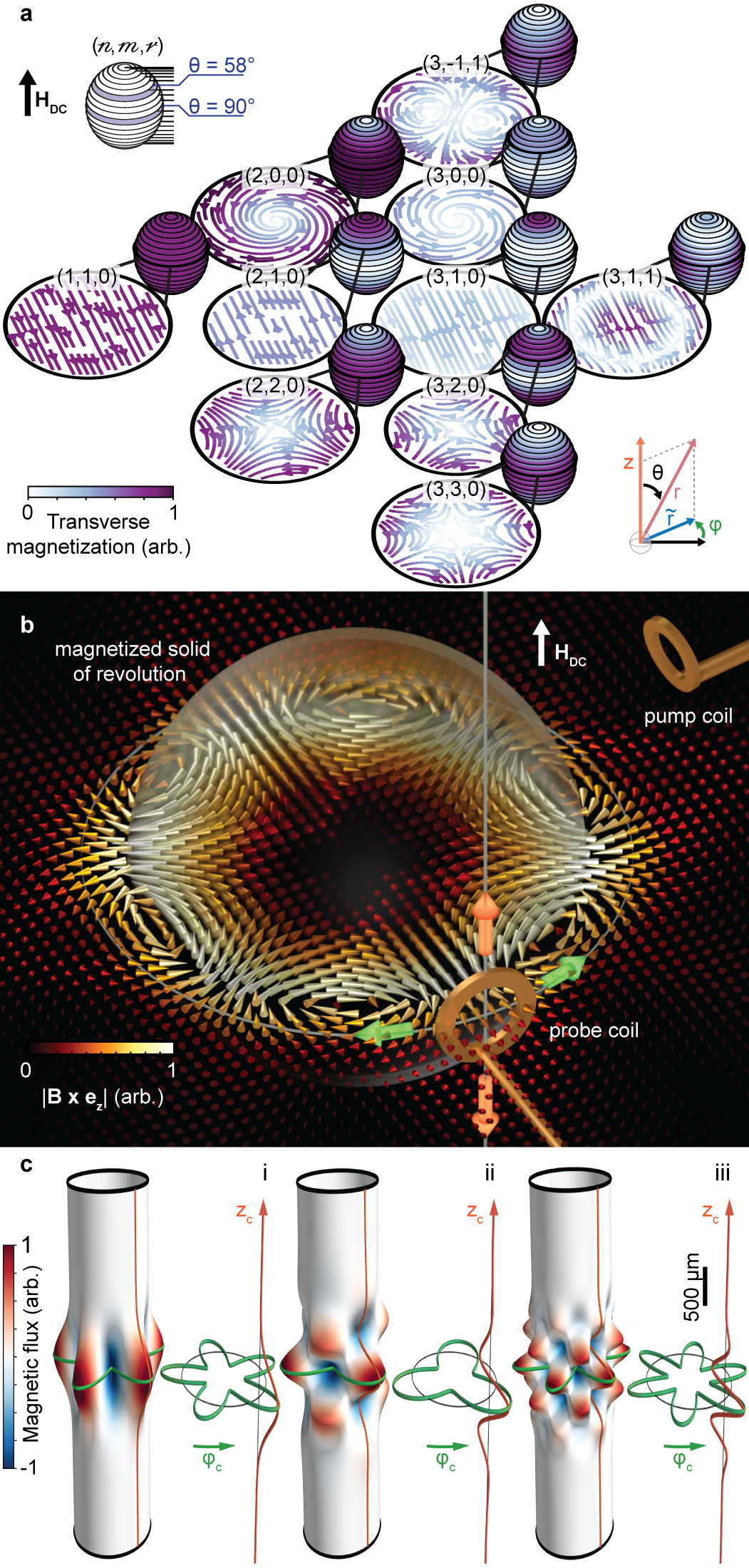}
\caption{\label{figure1} \textbf{Imaging the stray field induced by the spin-wave modes of a magnetized sphere.} 
\textbf{a,} Snapshot of the calculated spatial magnetization distribution of magnetostatic modes in a YIG sphere for $\mathcal{n} < 4$, transverse to the static magnetic field $H_\mathrm{DC}$. Their magnetization norm is color-encoded for successive sections from $\theta = 10^\circ$ to $170^\circ$~\cite{SI}. \textbf{b,}~Schematic of the experiment. A fixed loop coil~(pump coil) excites the magnons at microwave frequencies. These modes generate a dynamic magnetic field, represented here in the equatorial plane for the $(\mathcal{n},\mathcal{m},\mathcal{r})= (3,3,0)$ mode. The stray field spectrum is captured by a second loop coil~(probe coil) at different azimuth-altitude $(\varphi_c,z_c)$ positions. \textbf{c,} Numerical calculations of the spatially-resolved magnetic flux intercepted by a rectangular coil ($\Delta_y = 0.5\,$mm, $\Delta_z = 0.3\,$mm, $\tilde{r}_c = 2\,$mm) on a YIG sphere for $(\mathcal{n},\mathcal{m}) =(5,5)$, $(7,3)$ and $(12,6)$ families~(i-iii), encoded in the surface colors and corrugations.} 
\end{figure}	

\paragraph*{}
These spin-wave modes can be described within the magnetostatic approximation for a saturated magnetic ellipsoid~\cite{Walker1957} and their eigenfrequencies determined numerically. Diverse effects could alter this description: related to the environment as the temperature dependence of the saturation magnetization, the non-uniformity of the saturating static magnetic field or the presence of close-by parasitic elements~\cite{Dillon1958,Haigh2018}, or related to the sample as propagation corrections~\cite{FletcherPR1959}, magneto-crystalline anisotropy~\cite{Solt1960} or irregularities and composition defects, potentially resulting in inter-mode coupling~\cite{FletcherJAP1959b}. Increasing the optomagnonic coupling by using hybrid custom shapes with smaller mode volumes~\cite{Kusminskiy2016} will disturb the frequency distribution~\cite{Walker1958}. Associated with the density of modes excited by a non-uniform microwave field, the simple identification of the modes beyond the uniform precession mode using their expected ferromagnetic resonance frequencies~\cite{FletcherPR1959} is possibly ambiguous. 
An \textit{in situ} structural mapping of the spin-wave modes is needed to properly identify them and estimate their coupling to others modes, independently of the sample nature and of the experimental conditions, for fundamental magnonic studies~\cite{Klinger2017,Maier2017}, quantum magnonics~\cite{Tabuchi2015,Lachance-Quirion2017} and hybrid system operations~\cite{Kostylev2016,Zhang2016B,Osada2018,Osada2018B,Sharma2017,Sharma2018}. 
\paragraph*{}
Pushed forward by the demands for fast recording and high-capacity storage devices, magnetization dynamics of micro and nanostructures on a substrate have been intensively studied in the last years with advanced magnetic microscopy methods involving x-ray magnetic circular dichroism~\cite{Stoll2004}, magneto-optical interactions~\cite{Tamaru2002}, thermal effects~\cite{An2013}, microwave near-field~\cite{Lee2000} and magnetic scanning probe~\cite{Lee2010}.
At the other extreme, space exploration and geophysics have been implementing successfully magnetic field measurements around gigantic solids of revolution from satellite-based loop coil magnetometers employed to understand terrestrial polar aurorae from the ionosphere~\cite{Gurnett1964} to vector fluxgate magnetometers used to determine the magnetosphere and interior structures of Jupiter~\cite{Connerney2017b}. 
\paragraph*{}
Here, we access the spatial structure of the spin-wave modes of a bulk magnetized solid by measuring the magnetic flux spectrum intercepted by a mobile loop coil facing the sample at different azimuth-altitude positions, while the magnons are coherently excited by a fixed microwave antenna --- realizing a magnetic resonance imaging (MRI)~\cite{Lauterbur1973} scanner for collective electron spins excitations. 
The spatially and spectrally resolved magnetic responses of the system are carefully processed to extricate the nature of each spin-wave mode.

\paragraph*{}
A strong magnetic field $H_\mathrm{DC}$ applied along the $z$-axis saturates the magnetization of the sample. Solving Maxwell and Landau-Lifshitz equations reveals the existence of dynamic magnetization modes in the orthogonal plane~\cite{Walker1958}. For a millimeter-sized YIG sphere, these spin-wave modes have typical eigenfrequency $\Omega_k/2\pi \sim 5$--$10$\,GHz for $H_\mathrm{DC}$ in the $100\,$mT/$\mu_0$ range (with $\mu_0$ the magnetic constant).
In the magnetostatic approximation~\cite{FletcherJAP1959}, they are described by three indices $(\mathcal{n},\mathcal{m},\mathcal{r})$: $\mathcal{m}$ expresses the azimuthal dependency and is linked to the winding number of the spin texture~\cite{Osada2018}, $\mathcal{n}-|\mathcal{m}|$ the polar dependency and $\mathcal{r}$ the number of nodes along the radial direction. Their spatial distribution is deduced from a magnetic potential solution of Laplace equation in spheroidal coordinates dependent on the applied static magnetic field and on the considered mode. The non-trivial spatial distributions of the lowest-order modes are pictured in Fig.~\ref{figure1}\textbf{a}, their phase dependence at fixed altitude being given approximately by $-(\mathcal{m}-1)\varphi$. Outside the sphere, a magnetostatic mode induces a magnetic field $\mathbf{H}^\mathcal{m}_\mathcal{n} = \boldsymbol{\nabla} {\psi}^\mathcal{m}_\mathcal{n}$ such that close to the magnon resonance 
\begin{equation*}
\psi^\mathcal{m}_\mathcal{n}(r,\theta,\varphi) = \frac{\zeta^\mathcal{m}_\mathcal{n}}{r^{\mathcal{n}+1}} P^\mathcal{m}_\mathcal{n}(\cos \theta) e^{i \mathcal{m}\varphi}
\end{equation*}
with $\zeta^\mathcal{m}_\mathcal{n}$ encapsulating the pump field projection on the spin-wave mode and the resonance condition, while $P^\mathcal{m}_\mathcal{n}$ is the Ferrers function~\cite{SI}. The spherical coordinate system $(r,\theta,\varphi)$ is depicted in the inset of Fig.~\ref{figure1}\textbf{a}. 

\paragraph*{}
The imaging scheme is illustrated on Figure~\ref{figure1}\textbf{b}. The pump coil, fixed during the experiment, applies a microwave field exciting the magnons. The excitation frequency is swept to measure the spectral response of the whole magnonic system. The probe coil turns around the sample axis of revolution along a cylindrical orbit $(\varphi_c, z_c)$ and intercepts at each position the phase-resolved spectrum of the induced magnetic flux.
Working at a \textit{cylindrical} detection distance $\tilde{r}_c$ large compared to the typical width $2\Delta_y$ and height $2\Delta_z$ of the probe coil, the induced flux due to a $(\mathcal{n},\mathcal{m})$ mode can be approximated to $\phi^\mathcal{m}_\mathcal{n}(\varphi_c, z_c) \sim e^{i \mathcal{m} \varphi_c} \mathcal{Z}^\mathcal{m}_\mathcal{n}(z_c)$ with $\mathcal{Z}^\mathcal{m}_\mathcal{n}(z_c)$ having a mode-dependent envelope whose number of nodes along the altitude axis is related to $\mathcal{n}-|\mathcal{m}|$~\cite{SI}. 
Figure~\ref{figure1}\textbf{c} presents numerical computations of the magnetic flux induced by some representative mode families. 
Rotating the probe coil around the sample at fixed altitude grants access to the azimuthal parameter $\mathcal{m}$, enclosed in the mode relative phase, while a walk along the altitude $z$-axis leads to the polar parameter $\mathcal{n}-|\mathcal{m}|$. The radial dependency, not affecting the spatial distribution of the stray field, has to be deduced from the suite of eigenfrequencies of a given family $(\mathcal{n},\mathcal{m})$~\cite{Walker1958}. Each excited mode, with its distinct spectral signature, will contribute to the total stray field. The extraction of these features in the measured spectra at all positions along these two axes leads to their spatial mapping and subsequently to their robust identification.
\begin{figure}
\includegraphics[scale=0.95]{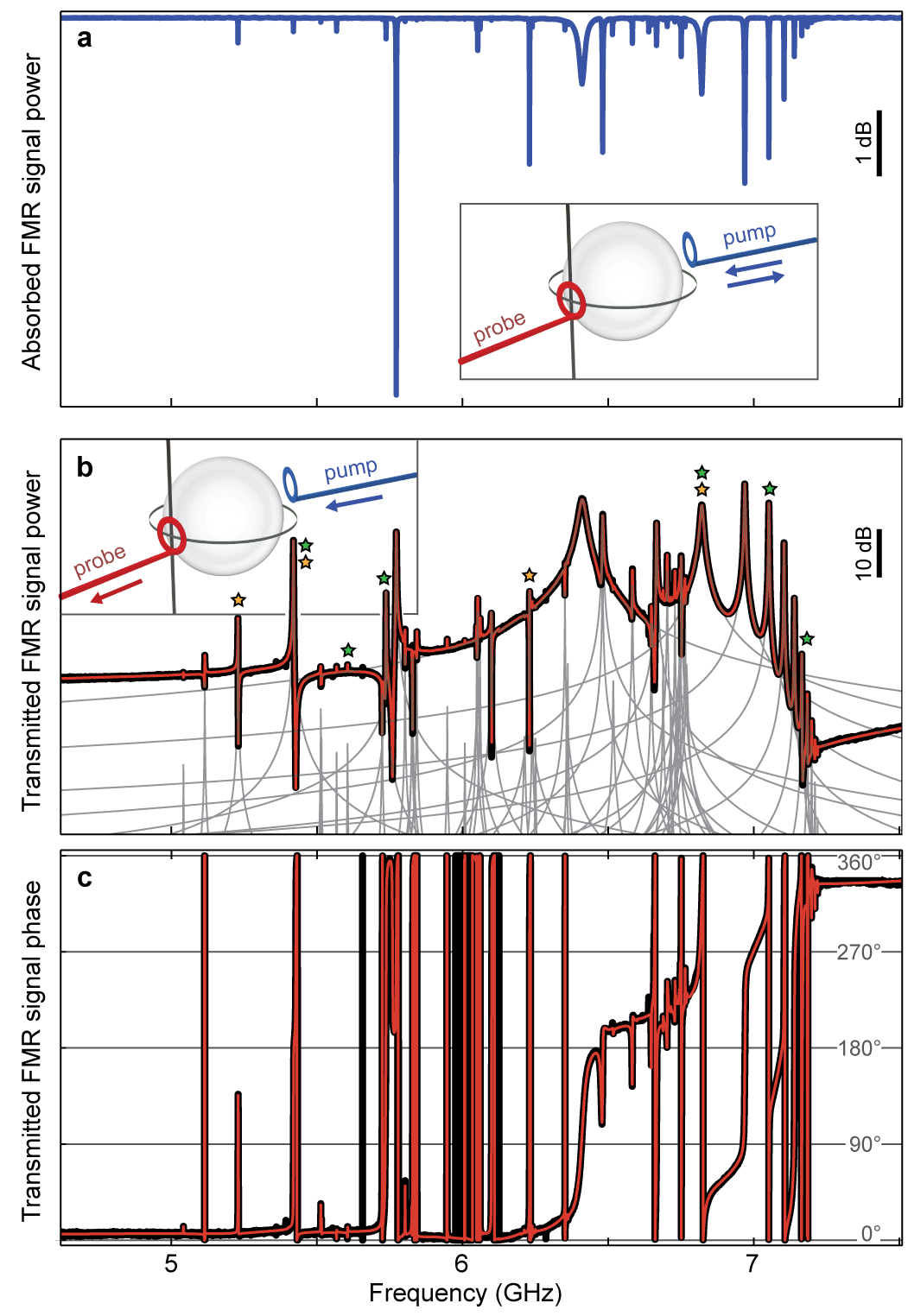}
\caption{\label{figure2}\textbf{Broadband ferromagnetic resonances (FMR) at a fixed azimuth-altitude coordinate.} \textbf{a,} Typical spectrum of the microwave power reflected by a $2$-mm YIG sphere into the pump coil (blue line). The absorption dips correspond to the resonance of more than 50 different individual spin-wave modes (resolution bandwidth: $10\,$kHz). The excitation microwave frequency range is chosen such that the spectrum encapsulates all the excited modes in one run. 
At a fixed probe-coil azimuth-altitude coordinate $(\varphi_c,z_c)$, the transmitted microwave spectra in power \textbf{(b)} and phase \textbf{(c)} (black line) intercepted by the probe coil through the sphere give access to the local information on each mode once fitted as a collection (red line) of individual harmonic oscillators (grey lines), in particular the local phase with respect to the origin fixed by the excitation field. The relative phase differences of the modes are responsible for non-trivial interference patterns. All the measurements are repeated after an automatized vertical retraction of the sample to define robust phase references~\cite{SI}. Colored stars mark the modes analyzed along the azimuth in Fig.~\ref{figure3} (green) and along the altitude in Fig.~\ref{figure4} (orange).} 
\end{figure}
\begin{figure*}
\includegraphics[scale=.95]{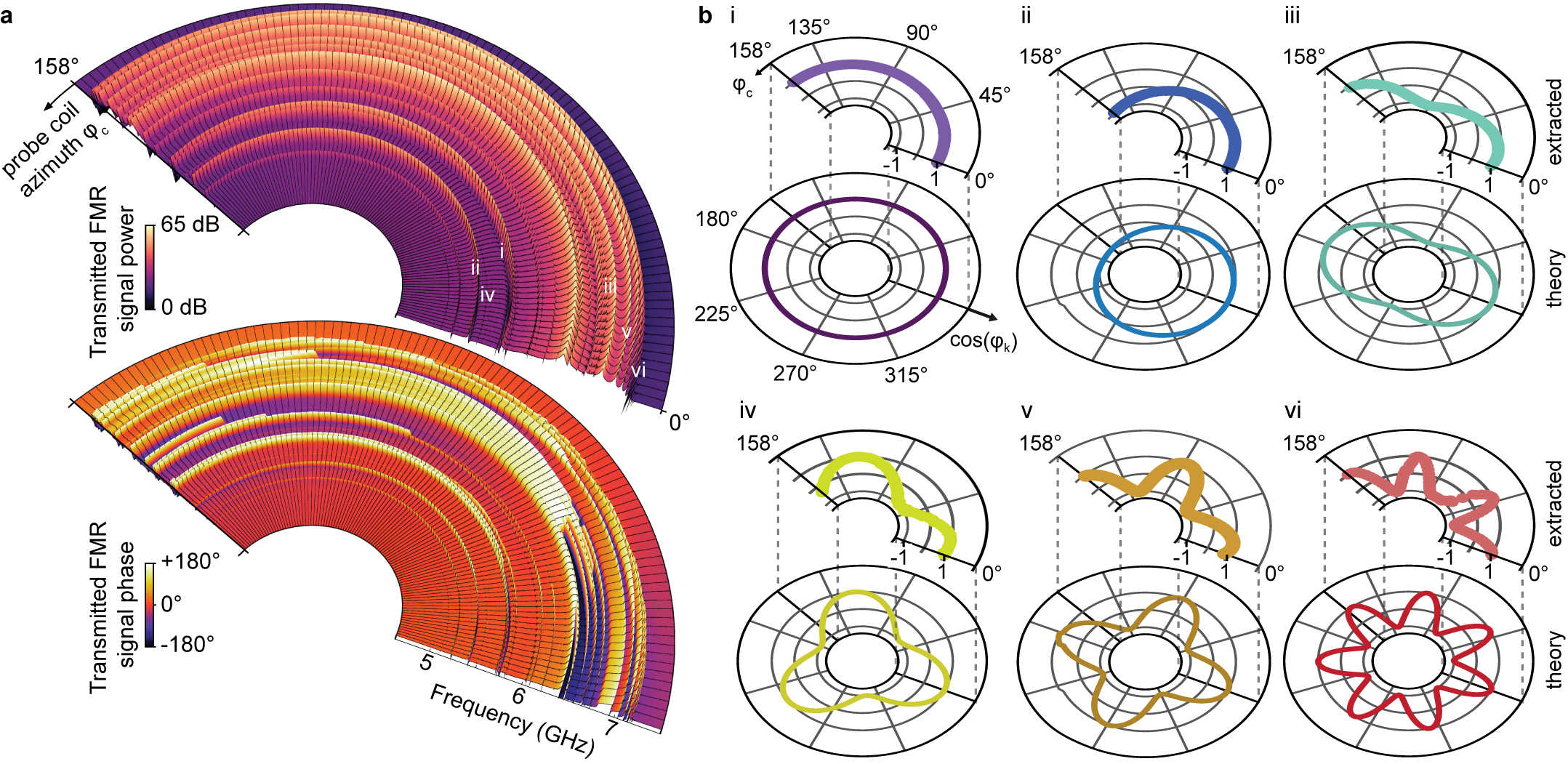}
\caption{\label{figure3} \textbf{Azimuthal dependence of the induced magnetic flux.} \textbf{a,} Microwave transmission spectra in power and phase as a function of the azimuthal position of the probe coil $\varphi_c$ in the sample equatorial plane. These spectra are all processed (see Figure~\ref{figure2} and Supplemental Material) to extract the relative phase of each spin-wave modes $\varphi_k$, exemplified by six of them in \textbf{b}(i-vi) (colored filled circles, top), with mean eigenfrequencies of $5.74\,$GHz, $5.42\,$GHz, $6.82\,$GHz, $5.61\,$GHz, $7.05\,$GHz and $7.17\,$GHz (indicated on \textbf{a} and on Fig.~\ref{figure2}\textbf{b} with green stars), exhibiting an azimuthal parameter $\mathcal{m}$ = 0, 1, 2, 3, 4 and 7 respectively, in comparison with the theoretical evolution of $\mathcal{m}\varphi_c$ (solid color lines, bottom). }
\end{figure*}
\begin{figure}
\includegraphics[scale=.9]{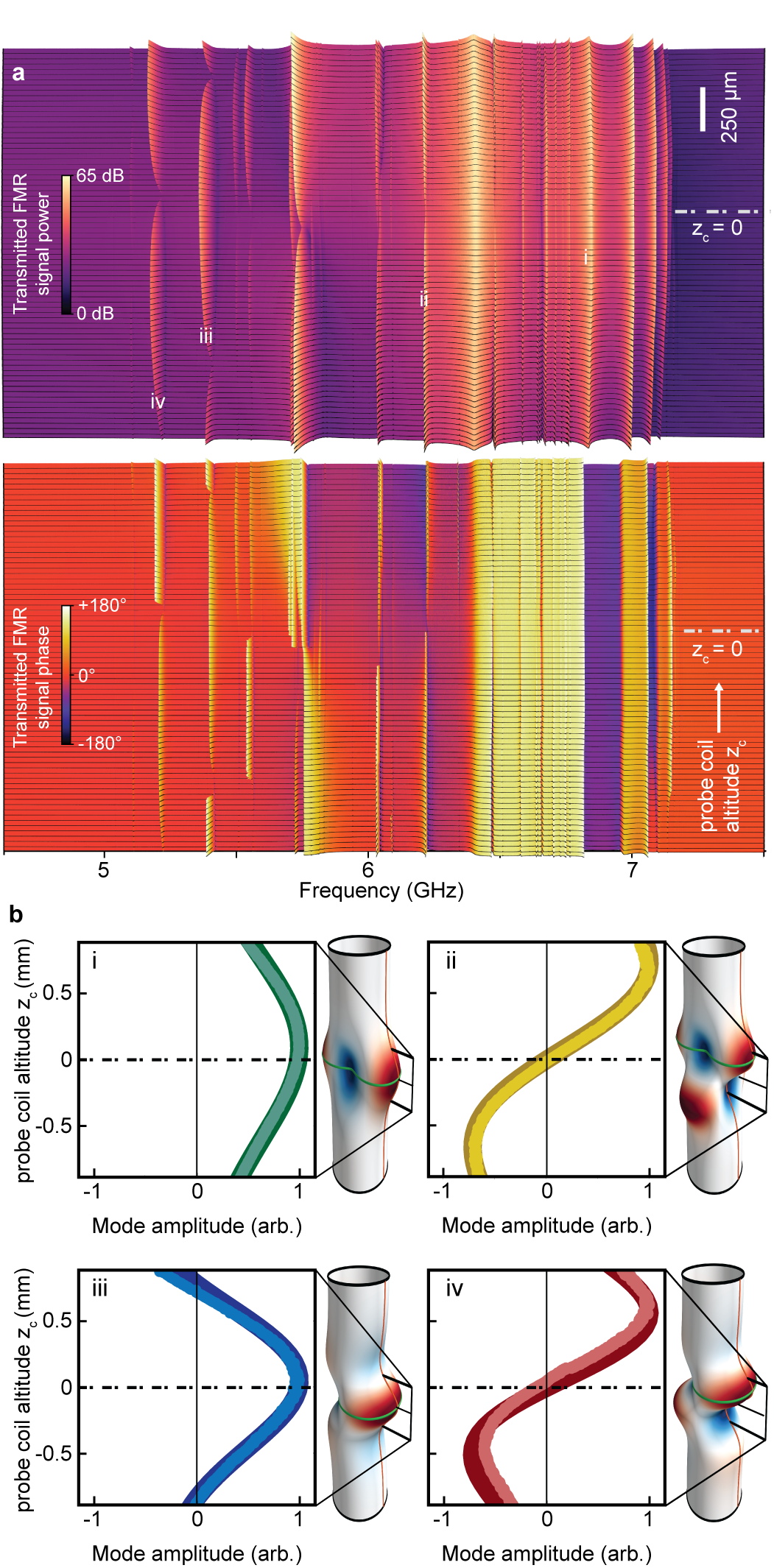}
\caption{\label{figure4}\textbf{Altitude dependence of the induced magnetic flux.} \textbf{a,} Microwave transmission spectra in power and phase as a function of the altitude position $z_c$ of the probe coil. \textbf{b,} Extraction of the local response of the modes allowing the reconstruction of their structure along the altitude axis (light-color filled circles), exemplified here for the lower-order families (i) $\mathcal{n}-|\mathcal{m}| = 0$, (ii) $\mathcal{n}-|\mathcal{m}| = 1$, (iii) $\mathcal{n}-|\mathcal{m}| = 2$ and (iv) $\mathcal{n}-|\mathcal{m}| = 3$. They are illustrated respectively by the modes of mean eigenfrequencies $6.82\,$GHz, $6.23\,$GHz, $5.42\,$GHz and $5.23\,$GHz (indicated on \textbf{a} and on Fig.~\ref{figure2}\textbf{b} with orange stars), identified as $(2,2,0)$, $(3,2,0)$, $(3,1,0)$ and $(4,1,0)$. Note that (i) and (iii) were represented in Fig.~\ref{figure3}, respectively labeled (iii) ($\mathcal{m}=2$) and (ii) ($\mathcal{m}=1$). The solid dark lines correspond to the expected flux ($\tilde{r}_c=2\,$mm, $\Delta_y=0.5\,$mm, $\Delta_z=0.3\,$mm) with a coil axis tilt $\xi_z=-6^{\circ}$. The altitude extent of these plots is depicted as a black rectangle in the insets showing the magnetic flux distribution.} 
\end{figure}
\paragraph*{}
The studied sample is a 2$\,$mm-diameter YIG sphere, placed at the center of an iron magnetic circuit ended by two permanent ring magnets distanced by 15$\, $mm. The static magnetic field ($\sim 230\,$mT/$\mu_0$) created along their revolution axis $z$ saturates the sphere along its [110] crystal axis.
The two small loop coils with a sub-millimetric inner radius ($\Delta_y \sim 0.5\,$mm, $\Delta_z \sim 0.3\,$mm), made out of semi-rigid coaxial copper cables by terminating their ends, are facing the sample and are respectively connected to the output (pump) and input (probe) ports of a vector network analyzer. The fixed pump coil stands a few millimeters away from the sample. Fixed to a three-axis linear actuator on a motorized rotation stage, the probe coil takes arbitrary positions $(\tilde{r}_c, \varphi_c, z_c)$ along a cylindrical orbit around the sample. The detection distance is set at $\tilde{r}_c = 2\,$mm. The reflected power from the pump coil (Fig.~\ref{figure2}\textbf{a}) reveals a collection of absorption dips, signatures of individual resonant spin-wave modes which can be modeled as damped harmonic oscillators~\cite{Gurevich1996}. We measure in transmission (Fig.~\ref{figure2}\textbf{b-c}) the magnetic flux intercepted by the probe coil at a particular position 
 \begin{math}S_\phi[\Omega]~=~\sum_{k = \{\mathcal{n},\mathcal{m},\mathcal{r}\}}{A_{k} e^{i\underline{\varphi}_{k}} / (\Omega^2 - \Omega_{k}^2 - i\Gamma_{k} \Omega)}\end{math} with $k$ running on all the excited modes, $\Omega_k / 2\pi$ the spin-wave mode eigenfrequency, $\Gamma_k$ its damping rate, and $A_k$ its relative response amplitude depending on the mode-dependent pump efficiency and on the mode spatial structure. The relative mode phase with respect to the excitation field, $\underline{\varphi}_{k} = \varphi^0_k + \varphi_k$, can be decomposed such that $\varphi^0_k$ is the mode phase origin defined by the excitation field and $\varphi_k$ its spatial component.
\paragraph*{}
For imaging the spatial structure of the spin-wave modes, we assemble these spectral measurements at numerous probe coordinates $(\varphi_c,z_c)$ and adjust the local spin-wave mode responses. First at fixed altitude, the coil travels around the sample.
Figure~\ref{figure3}\textbf{a} depicts the transmitted ferromagnetic resonance signal power and phase spectra along the azimuth in the equatorial plane ($z_c = 0$). While the modes amplitude is globally constant, their relative phase $\varphi_k$ individually changes with the probe azimuth. 
We report the evolution of the relative phase as a function of the coil azimuthal position. Figure~\ref{figure3}\textbf{b}(i-vi) illustrates the extracted azimuthal dependence of the phase of six different modes, with their theoretical counterparts appended, insuring a clear identification of modes with $\mathcal{m}=$ 0, 1, 2, 3, 4 and 7. 

\paragraph*{}
Next, at fixed azimuth we record the spectra along the altitude axis (Figure~\ref{figure4}\textbf{a}). As we are traveling over the mode envelope, the variation of the mode transmitted power and phase flips are observable directly on the spectra for well-isolated dominant modes. We report the extracted signed mode amplitude $A_k \cos \varphi_k$ as a function of the altitude of four representative modes in Fig.~\ref{figure4}\textbf{b}(i-iv), illustrating polar mode families $\mathcal{n}-|\mathcal{m}| = 0$, $\mathcal{n}-|\mathcal{m}| = 1$, $\mathcal{n}-|\mathcal{m}| = 2$ and $\mathcal{n}-|\mathcal{m}| = 3$. 
The altitude axis of the coil is slightly tilted by $\xi_z = -6^\circ$ with respect to the $z$-axis defined by the permanent magnets~\cite{SI}. This induces a slight imbalance in the measured flux, favoring positive altitudes. The measurements are in very good agreement with the theoretical calculation of the flux taking into account this correction. 
The exhibited modes could be identified respectively as $(2,2,0)$, $(3,2,0)$, $(3,1,0)$ and $(4,1,0)$.
Their relative eigenfrequencies spacing to the uniform precession mode $(\Omega_k - \Omega_{110})/2\pi$ are computed in the magnetostatic approximation for comparison~\cite{SI}. The discrepancies with the experimental observations, respectively $15\,\%$, $-36\,\%$, $-8\%$ and $-8\,\%$, most likely due to magneto-crystalline anisotropy and propagation effects, underline the necessity to access the mode spatial properties \textit{in situ} to avoid a misidentification between close-by modes.

\paragraph*{}
In the current conditions, our method reveals the polar mode families up to $\mathcal{n}-|\mathcal{m}| = 3$ and azimuthal families up to $\mathcal{m} = 7$. This range could be further expanded by designing the pump antenna to maximize the exciting efficiency of specific spin-wave modes of interest. 
Detection artifacts due to the probe coil height should appear only for modes with $\mathcal{n}-|\mathcal{m}| > 11$ and can be outclassed furthermore by tuning $\Delta_z/\tilde{r}_c$~\cite{SI}, largely overcoming the spin texture complexity traditionally under study.

\paragraph*{}
We have developed a new broadband resonant tomography scanning method to map the spatial structure of spin-wave modes hosted in a magnetized solid of revolution, providing a robust mode identification. 
Demonstrated here on a 2-mm YIG sphere saturated along its [110] axis, this approach straightforwardly extends to ellipsoids~\cite{Walker1957}, disks and rods~\cite{Dillon1960,Joseph1961}, provided that the spin-wave modes exhibit linewidths smaller than their typical frequency splitting~\cite{SI}. This versatile imaging method, which does not require to have a perfect knowledge on the sample and its environment,  will be particularly relevant for studying the magnetization dynamics of emergent magnetic materials and structures whose shape, crystallinity and composition could be challenging to control.
Hybrid magnonic operations well beyond the uniform precession mode can be envisioned, broadening the scope of quantum magnonics, magnomechanics and cavity optomagnonics towards the emergence of  macroscopic quantum devices.

\paragraph*{}
The authors thank A. Osada, R. Yamazaki and Y. Tabuchi for fruitful interactions. This work was supported by JSPS KAKENHI (Grant Nos. 16F16364 and 26220601) and by JST ERATO project (Grant No. JPMJER1601). A.G. is an Overseas researcher under Postdoctoral Fellowship of Japan Society for the Promotion of Science.


\begin{thebibliography}{41}%
\makeatletter
\providecommand \@ifxundefined [1]{%
 \@ifx{#1\undefined}
}%
\providecommand \@ifnum [1]{%
 \ifnum #1\expandafter \@firstoftwo
 \else \expandafter \@secondoftwo
 \fi
}%
\providecommand \@ifx [1]{%
 \ifx #1\expandafter \@firstoftwo
 \else \expandafter \@secondoftwo
 \fi
}%
\providecommand \natexlab [1]{#1}%
\providecommand \enquote  [1]{``#1''}%
\providecommand \bibnamefont  [1]{#1}%
\providecommand \bibfnamefont [1]{#1}%
\providecommand \citenamefont [1]{#1}%
\providecommand \href@noop [0]{\@secondoftwo}%
\providecommand \href [0]{\begingroup \@sanitize@url \@href}%
\providecommand \@href[1]{\@@startlink{#1}\@@href}%
\providecommand \@@href[1]{\endgroup#1\@@endlink}%
\providecommand \@sanitize@url [0]{\catcode `\\12\catcode `\$12\catcode
  `\&12\catcode `\#12\catcode `\^12\catcode `\_12\catcode `\%12\relax}%
\providecommand \@@startlink[1]{}%
\providecommand \@@endlink[0]{}%
\providecommand \url  [0]{\begingroup\@sanitize@url \@url }%
\providecommand \@url [1]{\endgroup\@href {#1}{\urlprefix }}%
\providecommand \urlprefix  [0]{URL }%
\providecommand \Eprint [0]{\href }%
\providecommand \doibase [0]{http://dx.doi.org/}%
\providecommand \selectlanguage [0]{\@gobble}%
\providecommand \bibinfo  [0]{\@secondoftwo}%
\providecommand \bibfield  [0]{\@secondoftwo}%
\providecommand \translation [1]{[#1]}%
\providecommand \BibitemOpen [0]{}%
\providecommand \bibitemStop [0]{}%
\providecommand \bibitemNoStop [0]{.\EOS\space}%
\providecommand \EOS [0]{\spacefactor3000\relax}%
\providecommand \BibitemShut  [1]{\csname bibitem#1\endcsname}%
\let\auto@bib@innerbib\@empty
\bibitem [{\citenamefont {Gurevich}\ and\ \citenamefont
  {Melkov}(1996)}]{Gurevich1996}%
  \BibitemOpen
  \bibfield  {author} {\bibinfo {author} {\bibfnamefont {A.}~\bibnamefont
  {Gurevich}}\ and\ \bibinfo {author} {\bibfnamefont {G.~A.}\ \bibnamefont
  {Melkov}},\ }\href@noop {} {\emph {\bibinfo {title} {Magnetization
  Oscillations and Waves}}}\ (\bibinfo  {publisher} {CRC Press},\ \bibinfo
  {year} {1996})\BibitemShut {NoStop}%
\bibitem [{\citenamefont {Stancil}\ and\ \citenamefont
  {Prabhakar}(2009)}]{Stancil2009}%
  \BibitemOpen
  \bibfield  {author} {\bibinfo {author} {\bibfnamefont {D.}~\bibnamefont
  {Stancil}}\ and\ \bibinfo {author} {\bibfnamefont {A.}~\bibnamefont
  {Prabhakar}},\ }\href@noop {} {\emph {\bibinfo {title} {Spin Waves: Theory
  and Applications}}}\ (\bibinfo  {publisher} {Springer},\ \bibinfo {year}
  {2009})\BibitemShut {NoStop}%
\bibitem [{\citenamefont {Kruglyak}\ \emph {et~al.}(2010)\citenamefont
  {Kruglyak}, \citenamefont {Demokritov},\ and\ \citenamefont
  {Grundler}}]{Kruglyak2010}%
  \BibitemOpen
  \bibfield  {author} {\bibinfo {author} {\bibfnamefont {V.}~\bibnamefont
  {Kruglyak}}, \bibinfo {author} {\bibfnamefont {S.}~\bibnamefont
  {Demokritov}}, \ and\ \bibinfo {author} {\bibfnamefont {D.}~\bibnamefont
  {Grundler}},\ }\href@noop {} {\bibfield  {journal} {\bibinfo  {journal}
  {Journal of Physics D: Applied Physics}\ }\textbf {\bibinfo {volume} {43}},\
  \bibinfo {pages} {264001} (\bibinfo {year} {2010})}\BibitemShut {NoStop}%
\bibitem [{\citenamefont {Tabuchi}\ \emph {et~al.}(2015)\citenamefont
  {Tabuchi}, \citenamefont {Ishino}, \citenamefont {Noguchi}, \citenamefont
  {Ishikawa}, \citenamefont {Yamazaki}, \citenamefont {Usami},\ and\
  \citenamefont {Nakamura}}]{Tabuchi2015}%
  \BibitemOpen
  \bibfield  {author} {\bibinfo {author} {\bibfnamefont {Y.}~\bibnamefont
  {Tabuchi}}, \bibinfo {author} {\bibfnamefont {S.}~\bibnamefont {Ishino}},
  \bibinfo {author} {\bibfnamefont {A.}~\bibnamefont {Noguchi}}, \bibinfo
  {author} {\bibfnamefont {T.}~\bibnamefont {Ishikawa}}, \bibinfo {author}
  {\bibfnamefont {R.}~\bibnamefont {Yamazaki}}, \bibinfo {author}
  {\bibfnamefont {K.}~\bibnamefont {Usami}}, \ and\ \bibinfo {author}
  {\bibfnamefont {Y.}~\bibnamefont {Nakamura}},\ }\href@noop {} {\bibfield
  {journal} {\bibinfo  {journal} {Science}\ }\textbf {\bibinfo {volume}
  {349}},\ \bibinfo {pages} {405} (\bibinfo {year} {2015})}\BibitemShut
  {NoStop}%
\bibitem [{\citenamefont {Hisatomi}\ \emph {et~al.}(2016)\citenamefont
  {Hisatomi}, \citenamefont {Osada}, \citenamefont {Tabuchi}, \citenamefont
  {Ishikawa}, \citenamefont {Noguchi}, \citenamefont {Yamazaki}, \citenamefont
  {Usami},\ and\ \citenamefont {Nakamura}}]{Hisatomi2016}%
  \BibitemOpen
  \bibfield  {author} {\bibinfo {author} {\bibfnamefont {R.}~\bibnamefont
  {Hisatomi}}, \bibinfo {author} {\bibfnamefont {A.}~\bibnamefont {Osada}},
  \bibinfo {author} {\bibfnamefont {Y.}~\bibnamefont {Tabuchi}}, \bibinfo
  {author} {\bibfnamefont {T.}~\bibnamefont {Ishikawa}}, \bibinfo {author}
  {\bibfnamefont {A.}~\bibnamefont {Noguchi}}, \bibinfo {author} {\bibfnamefont
  {R.}~\bibnamefont {Yamazaki}}, \bibinfo {author} {\bibfnamefont
  {K.}~\bibnamefont {Usami}}, \ and\ \bibinfo {author} {\bibfnamefont
  {Y.}~\bibnamefont {Nakamura}},\ }\href {\doibase 10.1103/PhysRevB.93.174427}
  {\bibfield  {journal} {\bibinfo  {journal} {Phys. Rev. B}\ }\textbf {\bibinfo
  {volume} {93}},\ \bibinfo {pages} {174427} (\bibinfo {year}
  {2016})}\BibitemShut {NoStop}%
\bibitem [{\citenamefont {Kimble}(2008)}]{Kimble2008}%
  \BibitemOpen
  \bibfield  {author} {\bibinfo {author} {\bibfnamefont {H.~J.}\ \bibnamefont
  {Kimble}},\ }\href@noop {} {\bibfield  {journal} {\bibinfo  {journal}
  {Nature}\ }\textbf {\bibinfo {volume} {453}},\ \bibinfo {pages} {1023}
  (\bibinfo {year} {2008})}\BibitemShut {NoStop}%
\bibitem [{\citenamefont {Wehner}\ \emph {et~al.}(2018)\citenamefont {Wehner},
  \citenamefont {Elkouss},\ and\ \citenamefont {Hanson}}]{Wehner2018}%
  \BibitemOpen
  \bibfield  {author} {\bibinfo {author} {\bibfnamefont {S.}~\bibnamefont
  {Wehner}}, \bibinfo {author} {\bibfnamefont {D.}~\bibnamefont {Elkouss}}, \
  and\ \bibinfo {author} {\bibfnamefont {R.}~\bibnamefont {Hanson}},\
  }\href@noop {} {\bibfield  {journal} {\bibinfo  {journal} {Science}\ }\textbf
  {\bibinfo {volume} {362}} (\bibinfo {year} {2018})}\BibitemShut {NoStop}%
\bibitem [{\citenamefont {Barzanjeh}\ \emph {et~al.}(2015)\citenamefont
  {Barzanjeh}, \citenamefont {Guha}, \citenamefont {Weedbrook}, \citenamefont
  {Vitali}, \citenamefont {Shapiro},\ and\ \citenamefont
  {Pirandola}}]{Barzanjeh2015}%
  \BibitemOpen
  \bibfield  {author} {\bibinfo {author} {\bibfnamefont {S.}~\bibnamefont
  {Barzanjeh}}, \bibinfo {author} {\bibfnamefont {S.}~\bibnamefont {Guha}},
  \bibinfo {author} {\bibfnamefont {C.}~\bibnamefont {Weedbrook}}, \bibinfo
  {author} {\bibfnamefont {D.}~\bibnamefont {Vitali}}, \bibinfo {author}
  {\bibfnamefont {J.}~\bibnamefont {Shapiro}}, \ and\ \bibinfo {author}
  {\bibfnamefont {S.}~\bibnamefont {Pirandola}},\ }\href {\doibase
  10.1103/PhysRevLett.114.080503} {\bibfield  {journal} {\bibinfo  {journal}
  {Phys. Rev. Lett.}\ }\textbf {\bibinfo {volume} {114}},\ \bibinfo {pages}
  {080503} (\bibinfo {year} {2015})}\BibitemShut {NoStop}%
\bibitem [{\citenamefont {Osada}\ \emph {et~al.}(2016)\citenamefont {Osada},
  \citenamefont {Hisatomi}, \citenamefont {Noguchi}, \citenamefont {Tabuchi},
  \citenamefont {Yamazaki}, \citenamefont {Usami}, \citenamefont {Sadgrove},
  \citenamefont {Yalla}, \citenamefont {Nomura},\ and\ \citenamefont
  {Nakamura}}]{Osada2016}%
  \BibitemOpen
  \bibfield  {author} {\bibinfo {author} {\bibfnamefont {A.}~\bibnamefont
  {Osada}}, \bibinfo {author} {\bibfnamefont {R.}~\bibnamefont {Hisatomi}},
  \bibinfo {author} {\bibfnamefont {A.}~\bibnamefont {Noguchi}}, \bibinfo
  {author} {\bibfnamefont {Y.}~\bibnamefont {Tabuchi}}, \bibinfo {author}
  {\bibfnamefont {R.}~\bibnamefont {Yamazaki}}, \bibinfo {author}
  {\bibfnamefont {K.}~\bibnamefont {Usami}}, \bibinfo {author} {\bibfnamefont
  {M.}~\bibnamefont {Sadgrove}}, \bibinfo {author} {\bibfnamefont
  {R.}~\bibnamefont {Yalla}}, \bibinfo {author} {\bibfnamefont
  {M.}~\bibnamefont {Nomura}}, \ and\ \bibinfo {author} {\bibfnamefont
  {Y.}~\bibnamefont {Nakamura}},\ }\href {\doibase
  10.1103/PhysRevLett.116.223601} {\bibfield  {journal} {\bibinfo  {journal}
  {Phys. Rev. Lett.}\ }\textbf {\bibinfo {volume} {116}},\ \bibinfo {pages}
  {223601} (\bibinfo {year} {2016})}\BibitemShut {NoStop}%
\bibitem [{\citenamefont {Haigh}\ \emph {et~al.}(2016)\citenamefont {Haigh},
  \citenamefont {Nunnenkamp}, \citenamefont {Ramsay},\ and\ \citenamefont
  {Ferguson}}]{Haigh2016}%
  \BibitemOpen
  \bibfield  {author} {\bibinfo {author} {\bibfnamefont {J.~A.}\ \bibnamefont
  {Haigh}}, \bibinfo {author} {\bibfnamefont {A.}~\bibnamefont {Nunnenkamp}},
  \bibinfo {author} {\bibfnamefont {A.~J.}\ \bibnamefont {Ramsay}}, \ and\
  \bibinfo {author} {\bibfnamefont {A.~J.}\ \bibnamefont {Ferguson}},\ }\href
  {\doibase 10.1103/PhysRevLett.117.133602} {\bibfield  {journal} {\bibinfo
  {journal} {Phys. Rev. Lett.}\ }\textbf {\bibinfo {volume} {117}},\ \bibinfo
  {pages} {133602} (\bibinfo {year} {2016})}\BibitemShut {NoStop}%
\bibitem [{\citenamefont {Zhang}\ \emph
  {et~al.}(2016{\natexlab{a}})\citenamefont {Zhang}, \citenamefont {Zhu},
  \citenamefont {Zou},\ and\ \citenamefont {Tang}}]{Zhang2016}%
  \BibitemOpen
  \bibfield  {author} {\bibinfo {author} {\bibfnamefont {X.}~\bibnamefont
  {Zhang}}, \bibinfo {author} {\bibfnamefont {N.}~\bibnamefont {Zhu}}, \bibinfo
  {author} {\bibfnamefont {C.-L.}\ \bibnamefont {Zou}}, \ and\ \bibinfo
  {author} {\bibfnamefont {H.~X.}\ \bibnamefont {Tang}},\ }\href@noop {}
  {\bibfield  {journal} {\bibinfo  {journal} {Phys. Rev. Lett.}\ }\textbf
  {\bibinfo {volume} {117}},\ \bibinfo {pages} {123605} (\bibinfo {year}
  {2016}{\natexlab{a}})}\BibitemShut {NoStop}%
\bibitem [{\citenamefont {Osada}\ \emph
  {et~al.}(2018{\natexlab{a}})\citenamefont {Osada}, \citenamefont {Gloppe},
  \citenamefont {Hisatomi}, \citenamefont {Noguchi}, \citenamefont {Yamazaki},
  \citenamefont {Nomura}, \citenamefont {Nakamura},\ and\ \citenamefont
  {Usami}}]{Osada2018}%
  \BibitemOpen
  \bibfield  {author} {\bibinfo {author} {\bibfnamefont {A.}~\bibnamefont
  {Osada}}, \bibinfo {author} {\bibfnamefont {A.}~\bibnamefont {Gloppe}},
  \bibinfo {author} {\bibfnamefont {R.}~\bibnamefont {Hisatomi}}, \bibinfo
  {author} {\bibfnamefont {A.}~\bibnamefont {Noguchi}}, \bibinfo {author}
  {\bibfnamefont {R.}~\bibnamefont {Yamazaki}}, \bibinfo {author}
  {\bibfnamefont {M.}~\bibnamefont {Nomura}}, \bibinfo {author} {\bibfnamefont
  {Y.}~\bibnamefont {Nakamura}}, \ and\ \bibinfo {author} {\bibfnamefont
  {K.}~\bibnamefont {Usami}},\ }\href {\doibase 10.1103/PhysRevLett.120.133602}
  {\bibfield  {journal} {\bibinfo  {journal} {Phys. Rev. Lett.}\ }\textbf
  {\bibinfo {volume} {120}},\ \bibinfo {pages} {133602} (\bibinfo {year}
  {2018}{\natexlab{a}})}\BibitemShut {NoStop}%
\bibitem [{\citenamefont {Haigh}\ \emph {et~al.}(2018)\citenamefont {Haigh},
  \citenamefont {Lambert}, \citenamefont {Sharma}, \citenamefont {Blanter},
  \citenamefont {Bauer},\ and\ \citenamefont {Ramsay}}]{Haigh2018}%
  \BibitemOpen
  \bibfield  {author} {\bibinfo {author} {\bibfnamefont {J.~A.}\ \bibnamefont
  {Haigh}}, \bibinfo {author} {\bibfnamefont {N.~J.}\ \bibnamefont {Lambert}},
  \bibinfo {author} {\bibfnamefont {S.}~\bibnamefont {Sharma}}, \bibinfo
  {author} {\bibfnamefont {Y.~M.}\ \bibnamefont {Blanter}}, \bibinfo {author}
  {\bibfnamefont {G.~E.~W.}\ \bibnamefont {Bauer}}, \ and\ \bibinfo {author}
  {\bibfnamefont {A.~J.}\ \bibnamefont {Ramsay}},\ }\href {\doibase
  10.1103/PhysRevB.97.214423} {\bibfield  {journal} {\bibinfo  {journal} {Phys.
  Rev. B}\ }\textbf {\bibinfo {volume} {97}},\ \bibinfo {pages} {214423}
  (\bibinfo {year} {2018})}\BibitemShut {NoStop}%
\bibitem [{\citenamefont {Lodahl}\ \emph {et~al.}(2017)\citenamefont {Lodahl},
  \citenamefont {Mahmoodian}, \citenamefont {Stobbe}, \citenamefont
  {Rauschenbeutel}, \citenamefont {Schneeweiss}, \citenamefont {Volz},
  \citenamefont {Pichler},\ and\ \citenamefont {Zoller}}]{Lodahl2017}%
  \BibitemOpen
  \bibfield  {author} {\bibinfo {author} {\bibfnamefont {P.}~\bibnamefont
  {Lodahl}}, \bibinfo {author} {\bibfnamefont {S.}~\bibnamefont {Mahmoodian}},
  \bibinfo {author} {\bibfnamefont {S.}~\bibnamefont {Stobbe}}, \bibinfo
  {author} {\bibfnamefont {A.}~\bibnamefont {Rauschenbeutel}}, \bibinfo
  {author} {\bibfnamefont {P.}~\bibnamefont {Schneeweiss}}, \bibinfo {author}
  {\bibfnamefont {J.}~\bibnamefont {Volz}}, \bibinfo {author} {\bibfnamefont
  {H.}~\bibnamefont {Pichler}}, \ and\ \bibinfo {author} {\bibfnamefont
  {P.}~\bibnamefont {Zoller}},\ }\href {http://dx.doi.org/10.1038/nature21037}
  {\bibfield  {journal} {\bibinfo  {journal} {Nature}\ }\textbf {\bibinfo
  {volume} {541}},\ \bibinfo {pages} {473} (\bibinfo {year}
  {2017})}\BibitemShut {NoStop}%
\bibitem [{\citenamefont {Sharma}\ \emph {et~al.}(2017)\citenamefont {Sharma},
  \citenamefont {Blanter},\ and\ \citenamefont {Bauer}}]{Sharma2017}%
  \BibitemOpen
  \bibfield  {author} {\bibinfo {author} {\bibfnamefont {S.}~\bibnamefont
  {Sharma}}, \bibinfo {author} {\bibfnamefont {Y.~M.}\ \bibnamefont {Blanter}},
  \ and\ \bibinfo {author} {\bibfnamefont {G.~E.~W.}\ \bibnamefont {Bauer}},\
  }\href {\doibase 10.1103/PhysRevB.96.094412} {\bibfield  {journal} {\bibinfo
  {journal} {Phys. Rev. B}\ }\textbf {\bibinfo {volume} {96}},\ \bibinfo
  {pages} {094412} (\bibinfo {year} {2017})}\BibitemShut {NoStop}%
\bibitem [{SI()}]{SI}%
  \BibitemOpen
  \href@noop {} {}\bibinfo {note} {See Supplemental Material}\BibitemShut
  {NoStop}%
\bibitem [{\citenamefont {Walker}(1957)}]{Walker1957}%
  \BibitemOpen
  \bibfield  {author} {\bibinfo {author} {\bibfnamefont {L.~R.}\ \bibnamefont
  {Walker}},\ }\href {\doibase 10.1103/PhysRev.105.390} {\bibfield  {journal}
  {\bibinfo  {journal} {Phys. Rev.}\ }\textbf {\bibinfo {volume} {105}},\
  \bibinfo {pages} {390} (\bibinfo {year} {1957})}\BibitemShut {NoStop}%
\bibitem [{\citenamefont {Dillon}(1958)}]{Dillon1958}%
  \BibitemOpen
  \bibfield  {author} {\bibinfo {author} {\bibfnamefont {J.~F.}\ \bibnamefont
  {Dillon}},\ }\href {\doibase 10.1103/PhysRev.112.59} {\bibfield  {journal}
  {\bibinfo  {journal} {Phys. Rev.}\ }\textbf {\bibinfo {volume} {112}},\
  \bibinfo {pages} {59} (\bibinfo {year} {1958})}\BibitemShut {NoStop}%
\bibitem [{\citenamefont {Fletcher}\ \emph {et~al.}(1959)\citenamefont
  {Fletcher}, \citenamefont {Solt},\ and\ \citenamefont
  {Bell}}]{FletcherPR1959}%
  \BibitemOpen
  \bibfield  {author} {\bibinfo {author} {\bibfnamefont {P.}~\bibnamefont
  {Fletcher}}, \bibinfo {author} {\bibfnamefont {I.~H.}\ \bibnamefont {Solt}},
  \ and\ \bibinfo {author} {\bibfnamefont {R.}~\bibnamefont {Bell}},\
  }\href@noop {} {\bibfield  {journal} {\bibinfo  {journal} {Phys. Rev.}\
  }\textbf {\bibinfo {volume} {114}},\ \bibinfo {pages} {739} (\bibinfo {year}
  {1959})}\BibitemShut {NoStop}%
\bibitem [{\citenamefont {Solt}\ and\ \citenamefont
  {Fletcher}(1960)}]{Solt1960}%
  \BibitemOpen
  \bibfield  {author} {\bibinfo {author} {\bibfnamefont {I.~H.~J.}\
  \bibnamefont {Solt}}\ and\ \bibinfo {author} {\bibfnamefont {P.~C.}\
  \bibnamefont {Fletcher}},\ }\href@noop {} {\bibfield  {journal} {\bibinfo
  {journal} {Journal of Applied Physics}\ }\textbf {\bibinfo {volume} {31}},\
  \bibinfo {pages} {S100} (\bibinfo {year} {1960})}\BibitemShut {NoStop}%
\bibitem [{\citenamefont {Fletcher}\ and\ \citenamefont
  {Solt}(1959)}]{FletcherJAP1959b}%
  \BibitemOpen
  \bibfield  {author} {\bibinfo {author} {\bibfnamefont {P.~C.}\ \bibnamefont
  {Fletcher}}\ and\ \bibinfo {author} {\bibfnamefont {I.~H.~J.}\ \bibnamefont
  {Solt}},\ }\href@noop {} {\bibfield  {journal} {\bibinfo  {journal} {Journal
  of Applied Physics}\ }\textbf {\bibinfo {volume} {30}},\ \bibinfo {pages}
  {S181} (\bibinfo {year} {1959})}\BibitemShut {NoStop}%
\bibitem [{\citenamefont {Viola-Kusminskiy}\ \emph {et~al.}(2016)\citenamefont
  {Viola-Kusminskiy}, \citenamefont {Tang},\ and\ \citenamefont
  {Marquardt}}]{Kusminskiy2016}%
  \BibitemOpen
  \bibfield  {author} {\bibinfo {author} {\bibfnamefont {S.}~\bibnamefont
  {Viola-Kusminskiy}}, \bibinfo {author} {\bibfnamefont {H.~X.}\ \bibnamefont
  {Tang}}, \ and\ \bibinfo {author} {\bibfnamefont {F.}~\bibnamefont
  {Marquardt}},\ }\href {\doibase 10.1103/PhysRevA.94.033821} {\bibfield
  {journal} {\bibinfo  {journal} {Phys. Rev. A}\ }\textbf {\bibinfo {volume}
  {94}},\ \bibinfo {pages} {033821} (\bibinfo {year} {2016})}\BibitemShut
  {NoStop}%
\bibitem [{\citenamefont {Walker}(1958)}]{Walker1958}%
  \BibitemOpen
  \bibfield  {author} {\bibinfo {author} {\bibfnamefont {L.~R.}\ \bibnamefont
  {Walker}},\ }\href@noop {} {\bibfield  {journal} {\bibinfo  {journal}
  {Journal of Applied Physics}\ }\textbf {\bibinfo {volume} {29}},\ \bibinfo
  {pages} {318} (\bibinfo {year} {1958})}\BibitemShut {NoStop}%
\bibitem [{\citenamefont {Klingler}\ \emph {et~al.}(2017)\citenamefont
  {Klingler}, \citenamefont {Maier-Flaig}, \citenamefont {Dubs}, \citenamefont
  {Surzhenko}, \citenamefont {Gross}, \citenamefont {Huebl}, \citenamefont
  {Goennenwein},\ and\ \citenamefont {Weiler}}]{Klinger2017}%
  \BibitemOpen
  \bibfield  {author} {\bibinfo {author} {\bibfnamefont {S.}~\bibnamefont
  {Klingler}}, \bibinfo {author} {\bibfnamefont {H.}~\bibnamefont
  {Maier-Flaig}}, \bibinfo {author} {\bibfnamefont {C.}~\bibnamefont {Dubs}},
  \bibinfo {author} {\bibfnamefont {O.}~\bibnamefont {Surzhenko}}, \bibinfo
  {author} {\bibfnamefont {R.}~\bibnamefont {Gross}}, \bibinfo {author}
  {\bibfnamefont {H.}~\bibnamefont {Huebl}}, \bibinfo {author} {\bibfnamefont
  {S.~T.~B.}\ \bibnamefont {Goennenwein}}, \ and\ \bibinfo {author}
  {\bibfnamefont {M.}~\bibnamefont {Weiler}},\ }\href@noop {} {\bibfield
  {journal} {\bibinfo  {journal} {Applied Physics Letters}\ }\textbf {\bibinfo
  {volume} {110}},\ \bibinfo {pages} {092409} (\bibinfo {year}
  {2017})}\BibitemShut {NoStop}%
\bibitem [{\citenamefont {Maier-Flaig}\ \emph {et~al.}(2017)\citenamefont
  {Maier-Flaig}, \citenamefont {Klingler}, \citenamefont {Dubs}, \citenamefont
  {Surzhenko}, \citenamefont {Gross}, \citenamefont {Weiler}, \citenamefont
  {Huebl},\ and\ \citenamefont {Goennenwein}}]{Maier2017}%
  \BibitemOpen
  \bibfield  {author} {\bibinfo {author} {\bibfnamefont {H.}~\bibnamefont
  {Maier-Flaig}}, \bibinfo {author} {\bibfnamefont {S.}~\bibnamefont
  {Klingler}}, \bibinfo {author} {\bibfnamefont {C.}~\bibnamefont {Dubs}},
  \bibinfo {author} {\bibfnamefont {O.}~\bibnamefont {Surzhenko}}, \bibinfo
  {author} {\bibfnamefont {R.}~\bibnamefont {Gross}}, \bibinfo {author}
  {\bibfnamefont {M.}~\bibnamefont {Weiler}}, \bibinfo {author} {\bibfnamefont
  {H.}~\bibnamefont {Huebl}}, \ and\ \bibinfo {author} {\bibfnamefont
  {S.~T.~B.}\ \bibnamefont {Goennenwein}},\ }\href {\doibase
  10.1103/PhysRevB.95.214423} {\bibfield  {journal} {\bibinfo  {journal} {Phys.
  Rev. B}\ }\textbf {\bibinfo {volume} {95}},\ \bibinfo {pages} {214423}
  (\bibinfo {year} {2017})}\BibitemShut {NoStop}%
\bibitem [{\citenamefont {Lachance-Quirion}\ \emph {et~al.}(2017)\citenamefont
  {Lachance-Quirion}, \citenamefont {Tabuchi}, \citenamefont {Ishino},
  \citenamefont {Noguchi}, \citenamefont {Ishikawa}, \citenamefont {Yamazaki},\
  and\ \citenamefont {Nakamura}}]{Lachance-Quirion2017}%
  \BibitemOpen
  \bibfield  {author} {\bibinfo {author} {\bibfnamefont {D.}~\bibnamefont
  {Lachance-Quirion}}, \bibinfo {author} {\bibfnamefont {Y.}~\bibnamefont
  {Tabuchi}}, \bibinfo {author} {\bibfnamefont {S.}~\bibnamefont {Ishino}},
  \bibinfo {author} {\bibfnamefont {A.}~\bibnamefont {Noguchi}}, \bibinfo
  {author} {\bibfnamefont {T.}~\bibnamefont {Ishikawa}}, \bibinfo {author}
  {\bibfnamefont {R.}~\bibnamefont {Yamazaki}}, \ and\ \bibinfo {author}
  {\bibfnamefont {Y.}~\bibnamefont {Nakamura}},\ }\href@noop {} {\bibfield
  {journal} {\bibinfo  {journal} {Science Advances}\ }\textbf {\bibinfo
  {volume} {3}} (\bibinfo {year} {2017})}\BibitemShut {NoStop}%
\bibitem [{\citenamefont {Kostylev}\ \emph {et~al.}(2016)\citenamefont
  {Kostylev}, \citenamefont {Goryachev},\ and\ \citenamefont
  {Tobar}}]{Kostylev2016}%
  \BibitemOpen
  \bibfield  {author} {\bibinfo {author} {\bibfnamefont {N.}~\bibnamefont
  {Kostylev}}, \bibinfo {author} {\bibfnamefont {M.}~\bibnamefont {Goryachev}},
  \ and\ \bibinfo {author} {\bibfnamefont {M.~E.}\ \bibnamefont {Tobar}},\
  }\href@noop {} {\bibfield  {journal} {\bibinfo  {journal} {Appl. Phys.
  Lett.}\ }\textbf {\bibinfo {volume} {108}},\ \bibinfo {pages} {062402}
  (\bibinfo {year} {2016})}\BibitemShut {NoStop}%
\bibitem [{\citenamefont {Zhang}\ \emph
  {et~al.}(2016{\natexlab{b}})\citenamefont {Zhang}, \citenamefont {Zou},
  \citenamefont {Jiang},\ and\ \citenamefont {Tang}}]{Zhang2016B}%
  \BibitemOpen
  \bibfield  {author} {\bibinfo {author} {\bibfnamefont {X.}~\bibnamefont
  {Zhang}}, \bibinfo {author} {\bibfnamefont {C.-L.}\ \bibnamefont {Zou}},
  \bibinfo {author} {\bibfnamefont {L.}~\bibnamefont {Jiang}}, \ and\ \bibinfo
  {author} {\bibfnamefont {H.}~\bibnamefont {Tang}},\ }\href@noop {} {\bibfield
   {journal} {\bibinfo  {journal} {Science Advances}\ }\textbf {\bibinfo
  {volume} {2}} (\bibinfo {year} {2016}{\natexlab{b}})}\BibitemShut {NoStop}%
\bibitem [{\citenamefont {Osada}\ \emph
  {et~al.}(2018{\natexlab{b}})\citenamefont {Osada}, \citenamefont {Gloppe},
  \citenamefont {Nakamura},\ and\ \citenamefont {Usami}}]{Osada2018B}%
  \BibitemOpen
  \bibfield  {author} {\bibinfo {author} {\bibfnamefont {A.}~\bibnamefont
  {Osada}}, \bibinfo {author} {\bibfnamefont {A.}~\bibnamefont {Gloppe}},
  \bibinfo {author} {\bibfnamefont {Y.}~\bibnamefont {Nakamura}}, \ and\
  \bibinfo {author} {\bibfnamefont {K.}~\bibnamefont {Usami}},\ }\href@noop {}
  {\bibfield  {journal} {\bibinfo  {journal} {New J. Phys.}\ }\textbf {\bibinfo
  {volume} {20}} (\bibinfo {year} {2018}{\natexlab{b}})}\BibitemShut {NoStop}%
\bibitem [{\citenamefont {Sharma}\ \emph {et~al.}(2018)\citenamefont {Sharma},
  \citenamefont {Blanter},\ and\ \citenamefont {Bauer}}]{Sharma2018}%
  \BibitemOpen
  \bibfield  {author} {\bibinfo {author} {\bibfnamefont {S.}~\bibnamefont
  {Sharma}}, \bibinfo {author} {\bibfnamefont {Y.}~\bibnamefont {Blanter}}, \
  and\ \bibinfo {author} {\bibfnamefont {G.}~\bibnamefont {Bauer}},\
  }\href@noop {} {\bibfield  {journal} {\bibinfo  {journal} {Phys. Rev. Lett.}\
  }\textbf {\bibinfo {volume} {121}},\ \bibinfo {pages} {087205} (\bibinfo
  {year} {2018})}\BibitemShut {NoStop}%
\bibitem [{\citenamefont {Stoll}\ \emph {et~al.}(2004)\citenamefont {Stoll},
  \citenamefont {Puzic}, \citenamefont {van Waeyenberge}, \citenamefont
  {Fischer}, \citenamefont {Raabe}, \citenamefont {Buess}, \citenamefont
  {Haug}, \citenamefont {Höllinger}, \citenamefont {Back}, \citenamefont
  {Weiss},\ and\ \citenamefont {Denbeaux}}]{Stoll2004}%
  \BibitemOpen
  \bibfield  {author} {\bibinfo {author} {\bibfnamefont {H.}~\bibnamefont
  {Stoll}}, \bibinfo {author} {\bibfnamefont {A.}~\bibnamefont {Puzic}},
  \bibinfo {author} {\bibfnamefont {B.}~\bibnamefont {van Waeyenberge}},
  \bibinfo {author} {\bibfnamefont {P.}~\bibnamefont {Fischer}}, \bibinfo
  {author} {\bibfnamefont {J.}~\bibnamefont {Raabe}}, \bibinfo {author}
  {\bibfnamefont {M.}~\bibnamefont {Buess}}, \bibinfo {author} {\bibfnamefont
  {T.}~\bibnamefont {Haug}}, \bibinfo {author} {\bibfnamefont {R.}~\bibnamefont
  {Hollinger}}, \bibinfo {author} {\bibfnamefont {C.}~\bibnamefont {Back}},
  \bibinfo {author} {\bibfnamefont {D.}~\bibnamefont {Weiss}}, \ and\ \bibinfo
  {author} {\bibfnamefont {G.}~\bibnamefont {Denbeaux}},\ }\href@noop {}
  {\bibfield  {journal} {\bibinfo  {journal} {Appl. Phys. Lett.}\ }\textbf
  {\bibinfo {volume} {84}},\ \bibinfo {pages} {3328} (\bibinfo {year}
  {2004})}\BibitemShut {NoStop}%
\bibitem [{\citenamefont {Tamaru}\ \emph {et~al.}(2002)\citenamefont {Tamaru},
  \citenamefont {Bain}, \citenamefont {van~de Veerdonk}, \citenamefont
  {Crawford}, \citenamefont {Covington},\ and\ \citenamefont
  {Kryder}}]{Tamaru2002}%
  \BibitemOpen
  \bibfield  {author} {\bibinfo {author} {\bibfnamefont {S.}~\bibnamefont
  {Tamaru}}, \bibinfo {author} {\bibfnamefont {J.~A.}\ \bibnamefont {Bain}},
  \bibinfo {author} {\bibfnamefont {R.~J.~M.}\ \bibnamefont {van~de Veerdonk}},
  \bibinfo {author} {\bibfnamefont {T.~M.}\ \bibnamefont {Crawford}}, \bibinfo
  {author} {\bibfnamefont {M.}~\bibnamefont {Covington}}, \ and\ \bibinfo
  {author} {\bibfnamefont {M.~H.}\ \bibnamefont {Kryder}},\ }\href@noop {}
  {\bibfield  {journal} {\bibinfo  {journal} {Journal of Applied Physics}\
  }\textbf {\bibinfo {volume} {91}},\ \bibinfo {pages} {8034} (\bibinfo {year}
  {2002})}\BibitemShut {NoStop}%
\bibitem [{\citenamefont {An}\ \emph {et~al.}(2013)\citenamefont {An},
  \citenamefont {Yamaguchi}, \citenamefont {Uchida},\ and\ \citenamefont
  {Saitoh}}]{An2013}%
  \BibitemOpen
  \bibfield  {author} {\bibinfo {author} {\bibfnamefont {T.}~\bibnamefont
  {An}}, \bibinfo {author} {\bibfnamefont {K.}~\bibnamefont {Yamaguchi}},
  \bibinfo {author} {\bibfnamefont {K.}~\bibnamefont {Uchida}}, \ and\ \bibinfo
  {author} {\bibfnamefont {E.}~\bibnamefont {Saitoh}},\ }\href@noop {}
  {\bibfield  {journal} {\bibinfo  {journal} {Appl. Phys. Lett.}\ }\textbf
  {\bibinfo {volume} {103}},\ \bibinfo {pages} {052410} (\bibinfo {year}
  {2013})}\BibitemShut {NoStop}%
\bibitem [{\citenamefont {Lee}\ \emph {et~al.}(2000)\citenamefont {Lee},
  \citenamefont {Vlahacos}, \citenamefont {Feenstra}, \citenamefont {Schwartz},
  \citenamefont {Steinhauer}, \citenamefont {Wellstood},\ and\ \citenamefont
  {Anlage}}]{Lee2000}%
  \BibitemOpen
  \bibfield  {author} {\bibinfo {author} {\bibfnamefont {S.}~\bibnamefont
  {Lee}}, \bibinfo {author} {\bibfnamefont {C.~P.}\ \bibnamefont {Vlahacos}},
  \bibinfo {author} {\bibfnamefont {B.~J.}\ \bibnamefont {Feenstra}}, \bibinfo
  {author} {\bibfnamefont {A.}~\bibnamefont {Schwartz}}, \bibinfo {author}
  {\bibfnamefont {D.~E.}\ \bibnamefont {Steinhauer}}, \bibinfo {author}
  {\bibfnamefont {F.~C.}\ \bibnamefont {Wellstood}}, \ and\ \bibinfo {author}
  {\bibfnamefont {S.~M.}\ \bibnamefont {Anlage}},\ }\href@noop {} {\bibfield
  {journal} {\bibinfo  {journal} {Appl. Phys. Lett.}\ }\textbf {\bibinfo
  {volume} {77}},\ \bibinfo {pages} {4404} (\bibinfo {year}
  {2000})}\BibitemShut {NoStop}%
\bibitem [{\citenamefont {Lee}\ \emph {et~al.}(2010)\citenamefont {Lee},
  \citenamefont {Obukhov}, \citenamefont {Xiang}, \citenamefont {Hauser},
  \citenamefont {Yang}, \citenamefont {Banerjee}, \citenamefont {Pelekhov},\
  and\ \citenamefont {Hammel}}]{Lee2010}%
  \BibitemOpen
  \bibfield  {author} {\bibinfo {author} {\bibfnamefont {I.}~\bibnamefont
  {Lee}}, \bibinfo {author} {\bibfnamefont {Y.}~\bibnamefont {Obukhov}},
  \bibinfo {author} {\bibfnamefont {G.}~\bibnamefont {Xiang}}, \bibinfo
  {author} {\bibfnamefont {A.}~\bibnamefont {Hauser}}, \bibinfo {author}
  {\bibfnamefont {F.}~\bibnamefont {Yang}}, \bibinfo {author} {\bibfnamefont
  {P.}~\bibnamefont {Banerjee}}, \bibinfo {author} {\bibfnamefont {D.~V.}\
  \bibnamefont {Pelekhov}}, \ and\ \bibinfo {author} {\bibfnamefont {P.~C.}\
  \bibnamefont {Hammel}},\ }\href@noop {} {\bibfield  {journal} {\bibinfo
  {journal} {Nature}\ }\textbf {\bibinfo {volume} {466}},\ \bibinfo {pages}
  {845} (\bibinfo {year} {2010})}\BibitemShut {NoStop}%
\bibitem [{\citenamefont {Gurnett}\ and\ \citenamefont
  {O'Brien}(1964)}]{Gurnett1964}%
  \BibitemOpen
  \bibfield  {author} {\bibinfo {author} {\bibfnamefont {D.~A.}\ \bibnamefont
  {Gurnett}}\ and\ \bibinfo {author} {\bibfnamefont {B.~J.}\ \bibnamefont
  {O'Brien}},\ }\href {\doibase 10.1029/JZ069i001p00065} {\bibfield  {journal}
  {\bibinfo  {journal} {Journal of Geophysical Research}\ }\textbf {\bibinfo
  {volume} {69}},\ \bibinfo {pages} {65} (\bibinfo {year} {1964})}\BibitemShut
  {NoStop}%
\bibitem [{\citenamefont {Connerney}\ \emph {et~al.}(2017)\citenamefont
  {Connerney} \emph {et~al.}}]{Connerney2017b}%
  \BibitemOpen
  \bibfield  {author} {\bibinfo {author} {\bibfnamefont {J.~E.~P.}\
  \bibnamefont {Connerney}} \emph {et~al.},\ }\href {\doibase
  10.1126/science.aam5928} {\bibfield  {journal} {\bibinfo  {journal}
  {Science}\ }\textbf {\bibinfo {volume} {356}},\ \bibinfo {pages} {826}
  (\bibinfo {year} {2017})}\BibitemShut {NoStop}%
\bibitem [{\citenamefont {Lauterbur}(1973)}]{Lauterbur1973}%
  \BibitemOpen
  \bibfield  {author} {\bibinfo {author} {\bibfnamefont {P.~C.}\ \bibnamefont
  {Lauterbur}},\ }\href@noop {} {\bibfield  {journal} {\bibinfo  {journal}
  {Nature}\ }\textbf {\bibinfo {volume} {242}},\ \bibinfo {pages} {190}
  (\bibinfo {year} {1973})}\BibitemShut {NoStop}%
\bibitem [{\citenamefont {Fletcher}\ and\ \citenamefont
  {Bell}(1959)}]{FletcherJAP1959}%
  \BibitemOpen
  \bibfield  {author} {\bibinfo {author} {\bibfnamefont {P.~C.}\ \bibnamefont
  {Fletcher}}\ and\ \bibinfo {author} {\bibfnamefont {R.~O.}\ \bibnamefont
  {Bell}},\ }\href@noop {} {\bibfield  {journal} {\bibinfo  {journal} {Journal
  of Applied Physics}\ }\textbf {\bibinfo {volume} {30}},\ \bibinfo {pages}
  {687} (\bibinfo {year} {1959})}\BibitemShut {NoStop}%
\bibitem [{\citenamefont {Dillon}(1960)}]{Dillon1960}%
  \BibitemOpen
  \bibfield  {author} {\bibinfo {author} {\bibfnamefont {J.~F.}\ \bibnamefont
  {Dillon}},\ }\href@noop {} {\bibfield  {journal} {\bibinfo  {journal}
  {Journal of Applied Physics}\ }\textbf {\bibinfo {volume} {31}},\ \bibinfo
  {pages} {1605} (\bibinfo {year} {1960})}\BibitemShut {NoStop}%
\bibitem [{\citenamefont {Joseph}\ and\ \citenamefont
  {Schlomann}(1961)}]{Joseph1961}%
  \BibitemOpen
  \bibfield  {author} {\bibinfo {author} {\bibfnamefont {R.~I.}\ \bibnamefont
  {Joseph}}\ and\ \bibinfo {author} {\bibfnamefont {E.}~\bibnamefont
  {Schlomann}},\ }\href@noop {} {\bibfield  {journal} {\bibinfo  {journal}
  {Journal of Applied Physics}\ }\textbf {\bibinfo {volume} {32}},\ \bibinfo
  {pages} {1001} (\bibinfo {year} {1961})}\BibitemShut {NoStop}%
\end{thebibliography}
%

\end{document}


\title{Supplemental Material\\ Resonant magnetic induction tomography of a magnetized sphere}

\author{A. Gloppe}
\email[]{arnaud.gloppe@qc.rcast.u-tokyo.ac.jp}
\affiliation{Research Center for Advanced Science and Technology (RCAST), The University of Tokyo, Meguro-ku, Tokyo 153-8904, Japan}
\author{R. Hisatomi}
\affiliation{Research Center for Advanced Science and Technology (RCAST), The University of Tokyo, Meguro-ku, Tokyo 153-8904, Japan}
\author{Y. Nakata}
\affiliation{Research Center for Advanced Science and Technology (RCAST), The University of Tokyo, Meguro-ku, Tokyo 153-8904, Japan}
\author{Y. Nakamura}
\affiliation{Research Center for Advanced Science and Technology (RCAST), The University of Tokyo, Meguro-ku, Tokyo 153-8904, Japan}
\affiliation{Center for Emergent Matter Science (CEMS), RIKEN, Wako, Saitama 351-0198, Japan}
\author{K. Usami}
\affiliation{Research Center for Advanced Science and Technology (RCAST), The University of Tokyo, Meguro-ku, Tokyo 153-8904, Japan}

\date{\today}

\begin{abstract}
\end{abstract}

\maketitle

\makeatletter
\def\l@subsection#1#2{}
\def\l@subsubsection#1#2{}
\makeatother
\makeatletter
\renewcommand*\l@section{\@dottedtocline{1}{1.5em}{2.3em}}
\makeatother
\tableofcontents


%

\clearpage

\section{Methods summary}
\subsubsection*{Spin-wave modes calculations}
\paragraph*{}
The spin-wave modes in Fig.~1\textbf{a} are plotted at time $t = 0.2 \times 2\pi/\Omega_k$, with a saturation magnetization $M_s = 194\,$mT/$\mu_0$, a static magnetic field $H_\mathrm{DC} = 315\,$mT/$\mu_0$ and a gyromagnetic ratio $\gamma/2\pi=28$\,GHz/T.
The resonance frequency $\Omega_k/2\pi$ of a given spin-wave mode is numerically determined by solving the resonance equation~(S19) in these conditions, a requirement to define properly the spheroidal coordinate system in which the internal magnetic potential has explicit solutions, computed for each colatitude $\theta$. Once interpolated on a regular Cartesian grid, the internal magnetic potential is numerically differentiated to obtain the internal magnetic field. 
The transverse magnetization $\mathbf{M}^\mathcal{m}_{\mathcal{n}}$ is obtained by linear combination of the internal magnetic field components with oblate factors, as functions of $H_\mathrm{DC}$ and $M_s$~\cite{FletcherJAP1959}. The stray field is numerically evaluated from the external magnetic potential $\psi^\mathcal{m}_{\mathcal{n}}$. The internal and external fields for $(3,3,0)$ are joined to plot the total transverse magnetic induction field $\mathbf{B}^3_3$ in Figure~1\textbf{b}, slightly out of resonance for a better visualization.

\subsubsection*{Probe coil}
\paragraph*{}
The coil parameters $\Delta_y$, $\Delta_z$ and $\tilde{r}_c$ are chosen carefully to ensure proper imaging.
The detection distance $\tilde{r}_c$ should be short to maximize the acquired signal ($\phi^\mathcal{m}_\mathcal{n} \propto 1/\tilde{r}_c^{n+2}$) while insuring that $\Delta_y/\tilde{r}_c \ll 1$ and $\Delta_z/\tilde{r}_c \ll 1$. The lateral semi-extension $\Delta_y$ while small compared to $\tilde{r}_c$, should be maximized $(\phi^\mathcal{m}_\mathcal{n} \propto \Delta_y)$. The value of $\Delta_z$ should be small enough to guarantee the measured structure can be safely related to $\mathcal{n}-|\mathcal{m}|$. The choice of $\Delta_z = 0.3$\,mm for $\tilde{r}_c=2\,$mm insures the safe detection of spin-wave modes with $\mathcal{n}-|\mathcal{m}|$ up to 11. The distance $\tilde{r}_c$ may be slightly tuned to maximize the detection of a particular azimuth family (see Sec.~\ref{section:theo} for details).

\subsubsection*{Data acquisition}
\paragraph*{}
In the magnetostatic approximation~\cite{Walker1958}, the spin-wave modes are expected on a frequency range of $\gamma\mu_0 M_s/4\pi \sim 2.6\,$GHz. 
The information on all the observable spin-wave modes is obtained with a high resolution by acquiring the spectra by pieces, for a total range of 2.89$\,$GHz and more than 65,000 points by spectrum (microwave excitation power: $0$\,dBm).
At fixed altitude, we record spectra at 101 different probe coil azimuthal positions on a 158$^\circ$ range. The accessible angular range is limited by the presence of the magnetic circuit and pump coil. At fixed azimuth, we record spectra at 297 different probe altitudes on a $2.2\,$mm extent across the equatorial plane of the sphere. A residual background from the direct coupling between the coils and the parasitic response of close-by elements subsists with typical linewidths ($>1\,$GHz) much larger than those of the spin-wave modes ($<10\,$MHz) and could be then conscientiously dismissed.
A set of measurement at a given coil position lasts for 4$\,$min with a resolution bandwidth of 10$\,$kHz, resulting in a total measurement time of 6$\,$h along the azimuth (Figure~3) and 22$\,$h along the altitude (Figure~4).

\section{Magnetic flux induced by a magnetostatic mode of a ferromagnetic sphere}
\label{section:theo}
\begin{figure*}
\includegraphics[scale=1]{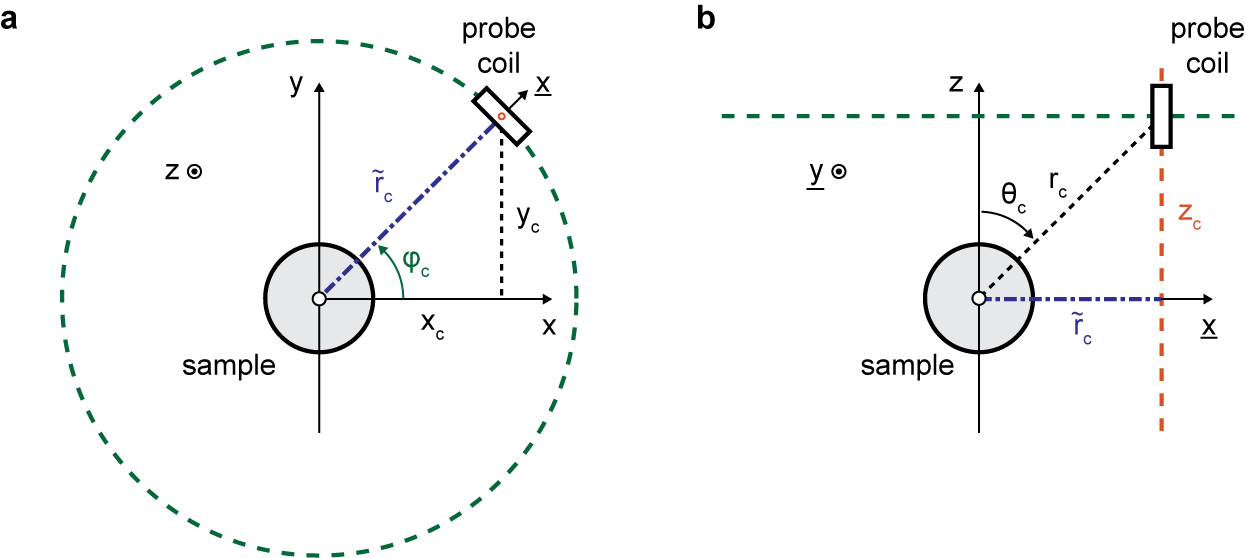}
\caption{\label{fig:repere} Frame of the problem figuring the position and orientation of the coil compared to the sample, in the ideal case in which the coil faces perfectly the sphere and its trajectory shares the same center and rotation axis, projected onto the $xy$-plane (\textbf{a}) and onto the $\underline{x}z$-plane rotating with $\varphi_c$ (\textbf{b}). The coil cylindrical trajectory is represented in color dashed lines, with the azimuth and altitude in green	and orange respectively. The horizontal distance of the probe coil $\tilde{r}_c$ is pictured in blue.}
\end{figure*}

\paragraph*{}
We compute the magnetic flux induced by a magnetostatic mode $(\mathcal{n},\mathcal{m})$ within a ferromagnetic sphere, intercepted by a coil facing the sample along a cylindrical orbit $(\tilde{r}_c,\varphi_c, z_c)$, represented in Figure~\ref{fig:repere}.

The magnetic potential outside the sample induced by the magnetostatic mode $(\mathcal{n},\mathcal{m})$, excited by a pump field, is given in spherical coordinates by~\cite{FletcherJAP1959}:
\begin{equation}
\psi^\mathcal{m}_\mathcal{n}(r,\theta,\varphi) = A^\mathcal{m}_\mathcal{n} r^\mathcal{n}\left[1 + \alpha^\mathcal{m}_\mathcal{n} \left(\frac{a}{r}\right)^{2\mathcal{n}+1} \right] P^\mathcal{m}_\mathcal{n}(\cos \theta) e^{i \mathcal{m}\varphi}\mathrm{,}
\end{equation}
with $P^\mathcal{m}_\mathcal{n}$ the Ferrers function \cite{Olver2010}, $a$ the sphere radius, $A^\mathcal{m}_\mathcal{n}$ related to the projection of the pump field on the spherical harmonics and $\alpha^\mathcal{m}_\mathcal{n}$ is an amplification factor related to the mode resonance. 

The gradient of this outer magnetic potential is 
\begin{equation*}
\boldsymbol{\nabla} \psi (r,\theta,\varphi) = \partial_r \psi \, \mathbf{e_r} + \frac{1}{r} \partial_\theta \psi \, \mathbf{e_\theta} + \frac{1}{r\sin\theta} \partial_\varphi \psi \, \mathbf{e_\varphi}\mathrm{,} 
\end{equation*}
with 
\begin{align*}
\partial_r \psi (r,\theta, \varphi) &= e^{i \mathcal{m} \varphi} \Delta_1(r) \, P^\mathcal{m}_\mathcal{n}(\cos \theta)\mathrm{,} \\
\partial_\theta \psi (r,\theta, \varphi) &= e^{i \mathcal{m} \varphi} \Delta_2(r) \, \frac{1}{\sin \theta} \left[-(\mathcal{n}+1) \cos \theta P^\mathcal{m}_\mathcal{n}(\cos \theta) + (\mathcal{n}+1-\mathcal{m}) P^\mathcal{m}_{\mathcal{n}+1}(\cos \theta) \right]\mathrm{,} \\
\partial_\varphi \psi (r,\theta, \varphi) &= e^{i \mathcal{m} \varphi} \Delta_2(r) \, i\mathcal{m} P^\mathcal{m}_\mathcal{n}(\cos \theta) \mathrm{,}
\end{align*}
and 
\begin{align}
\Delta_1(r) &\equiv A^\mathcal{m}_\mathcal{n} r^{\mathcal{n}-1} \left[\mathcal{n} - (\mathcal{n}+1) \, \alpha^\mathcal{m}_\mathcal{n} \left(\frac{a}{r}\right)^{2\mathcal{n}+1} \right] \mathrm{,}\\
\Delta_2(r)/r \equiv \tilde{\Delta}_2(r) &\equiv A^\mathcal{m}_\mathcal{n} r^{\mathcal{n}-1} \left[1 + \alpha^\mathcal{m}_\mathcal{n} \left(\frac{a}{r}\right)^{2\mathcal{n}+1} \right]\mathrm{.}
\end{align}
and will be written in the Cartesian basis $\mathcal{B}$ as
\begin{align*}
\boldsymbol{\nabla} \psi(r,\theta,\varphi) =
	&\left[\left( \partial_r \psi \, \sin \theta + \frac{1}{r}\partial_\theta \psi \, \cos \theta \right) \cos \varphi - \frac{1}{r\sin \theta}\partial_\varphi \psi \, \sin \varphi \right] \mathbf{e_x} \\ +
	&\left[\left(\partial_r \psi \, \sin \theta + \frac{1}{r}\partial_\theta \psi \, \cos \theta \right) \sin \varphi + \frac{1}{r\sin \theta} \partial_\varphi \psi \, \cos \varphi \right] \mathbf{e_y}	\\ + 
	&\left[\partial_r \psi \, \cos \theta - \frac{1}{r}\partial_\theta \psi \, \sin \theta \right] \mathbf{e_z} \mathrm{.}
\end{align*}
In the rotating frame of the probe coil $\mathcal{\underline{B}}$ such that $\begin{pmatrix}x\\y\\z\end{pmatrix}_\mathcal{B} = R_{\varphi_c}^{-1} \cdot \begin{pmatrix}\underline{x} \\ \underline{y} \\ \underline{z}\end{pmatrix}_\mathcal{\underline{B}}$ with the rotation matrix
\begin{equation*}
R_{\varphi_c} \equiv 
\begin{pmatrix}
\cos \varphi_c & \sin \varphi_c & 0 \\
-\sin \varphi_c & \cos \varphi_c & 0 \\
0 & 0 & 1 
\end{pmatrix}\mathrm{.}
\end{equation*}
For a perfectly oriented coil, the surface normal of the probe can be simply expressed as 
\begin{equation}
\label{eq:projectiondS}
R_{\varphi_c} \cdot \mathbf{dS}_\mathcal{B} = \mathbf{dS}_
\mathcal{\underline{B}} = dS \, \begin{pmatrix}1 \\ 0 \\ 0\end{pmatrix}_\mathcal{\underline{B}}
\end{equation}
and we can write the induced flux
\begin{equation}
\label{phimn}
\phi^\mathcal{m}_\mathcal{n} = \mu_0 \iint_S \, \boldsymbol{\nabla} \psi(r,\theta,\varphi) \cdot \mathbf{dS}_\mathcal{B} = \mu_0 \iint_S \, dS \left(I_r + I_\theta + I_\varphi \right)
\end{equation}
with 
\begin{align*}
I_r(r,\theta, \underline{\varphi}) &\equiv \sin \theta \cos \underline{\varphi} \, \partial_r \psi \mathrm{,}\\
I_\theta(r,\theta, \underline{\varphi}) &\equiv \frac{1}{r} \cos \theta \cos \underline{\varphi} \, \partial_\theta \psi \mathrm{,}\\
I_\varphi(r,\theta, \underline{\varphi}) &\equiv -\frac{1}{r} \frac{1}{\sin \theta} \sin \underline{\varphi} \, \partial_\varphi \psi \mathrm{,}
\end{align*}
defining $\underline{\varphi} \equiv \varphi - \varphi_c$. By developing these formulae we obtain, 
\begin{align}
I_r(r,\theta, \underline{\varphi}) &= e^{i \mathcal{m} \varphi_c} \,\,
\left[e^{i(\mathcal{m}+1)\underline{\varphi}} + e^{i(\mathcal{m}-1)\underline{\varphi}} \right]
\,\, \sin\theta \, P^\mathcal{m}_\mathcal{n}(\cos \theta) \, \frac{\Delta_1(r)}{2} \mathrm{,}
\\
I_\theta(r,\theta, \underline{\varphi}) &= e^{i \mathcal{m} \varphi_c} \,\,
\left[e^{i(\mathcal{m}+1)\underline{\varphi}} + e^{i(\mathcal{m}-1)\underline{\varphi}} \right]
\,\, \dfrac{ \cos \theta }{ \sin \theta}
 \left[-(\mathcal{n}+1) \cos\theta \, P^\mathcal{m}_\mathcal{n}(\cos \theta) + (\mathcal{n}+1-\mathcal{m}) P^\mathcal{m}_{\mathcal{n}+1}(\cos \theta) \right] \, \frac{\tilde{\Delta}_2(r)}{2} \mathrm{,}
\\
I_\varphi(r,\theta, \underline{\varphi}) &=  - e^{i \mathcal{m} \varphi_c} \,\,
\left[e^{i(\mathcal{m}+1)\underline{\varphi}} - e^{i(\mathcal{m}-1)\underline{\varphi}} \right]
\,\, \mathcal{m} \dfrac{1}{ \sin \theta} P^\mathcal{m}_\mathcal{n}(\cos \theta) \, \frac{\tilde{\Delta}_2(r)}{2} \mathrm{.}
\end{align}
These expressions can be computed numerically for an arbitrary coil section.
\subsection{A small rectangular coil}
\paragraph*{}
We consider now that the coil has a rectangular section with a lateral extension $2\Delta_y$ much smaller than the \textit{cylindrical} detection distance $\tilde{r}_c$, then 
\begin{equation*}
\underline{\varphi} = \arctan \left(\frac{\underline{y}}{\tilde{r}_c} \right) \sim \frac{\underline{y}}{\tilde{r}_c}
\end{equation*}
and the contribution of $\underline{y}$ to $r$ is negligible. 
Then Eq.(\ref{phimn}) could be then rewritten
\begin{equation*}
\phi^\mathcal{m}_\mathcal{n} = \mu_0 \int_{z_c-\Delta_z^-}^{z_c+\Delta_z^+}dz \int_{-\Delta_y^- }^{\Delta_y^+ } \, d\underline{y} \, \left(I_r + I_\theta + I_\varphi \right) \mathrm{.}
\end{equation*}
We define the integrals along the $\underline{y}$-direction

\begin{equation}
\mathcal{Y}^\pm \equiv \int_{-\Delta_y^- }^{\Delta_y^+ } d\underline{y} \, 
\left[e^{i(\mathcal{m}+1)\underline{\varphi}} \pm e^{i(\mathcal{m}-1)\underline{\varphi}} \right]
\end{equation}

and along the $z$-direction
\begin{align}
\label{eq:Zr}
\mathcal{Z}_r (z_c) &\equiv \int_{z_c-\Delta_z^-}^{z_c+\Delta_z^+}dz \, \sin(\theta) P^\mathcal{m}_\mathcal{n}(\cos \theta) \, \frac{\Delta_1(r)}{2} \mathrm{,}\\
\label{eq:Ztheta}
\mathcal{Z}_\theta (z_c) &\equiv \int_{z_c-\Delta_z^-}^{z_c+\Delta_z^+}dz\, \dfrac{ \cos \theta }{ \sin \theta}
 \left[-(\mathcal{n}+1) \cos\theta \, P^\mathcal{m}_\mathcal{n}(\cos \theta) + (\mathcal{n}+1-\mathcal{m}) P^\mathcal{m}_{\mathcal{n}+1}(\cos \theta) \right] \, \frac{\tilde{\Delta}_2(r)}{2} \mathrm{,} \\
\label{eq:Zphi}
\mathcal{Z}_\varphi (z_c) &\equiv \int_{z_c-\Delta_z^-}^{z_c+\Delta_z^+}dz \, \mathcal{m}\dfrac{1}{ \sin \theta} P^\mathcal{m}_\mathcal{n}(\cos \theta) \, \frac{\tilde{\Delta}_2(r)}{2} \mathrm{,} 
\end{align}
so that 
\begin{equation}
\label{eq:flux}
\phi^\mathcal{m}_\mathcal{n} = \mu_0 e^{i \mathcal{m} \varphi_c} \left[\mathcal{Y}^+ \left[ \mathcal{Z}_r(z_c) + \mathcal{Z}_\theta(z_c)\right] - \mathcal{Y}^- \,\, \mathcal{Z}_\varphi(z_c) \right]\mathrm{.}
\end{equation}

\subsection{Influence of the lateral extension}
\begin{figure*}[t]
\includegraphics[scale=1]{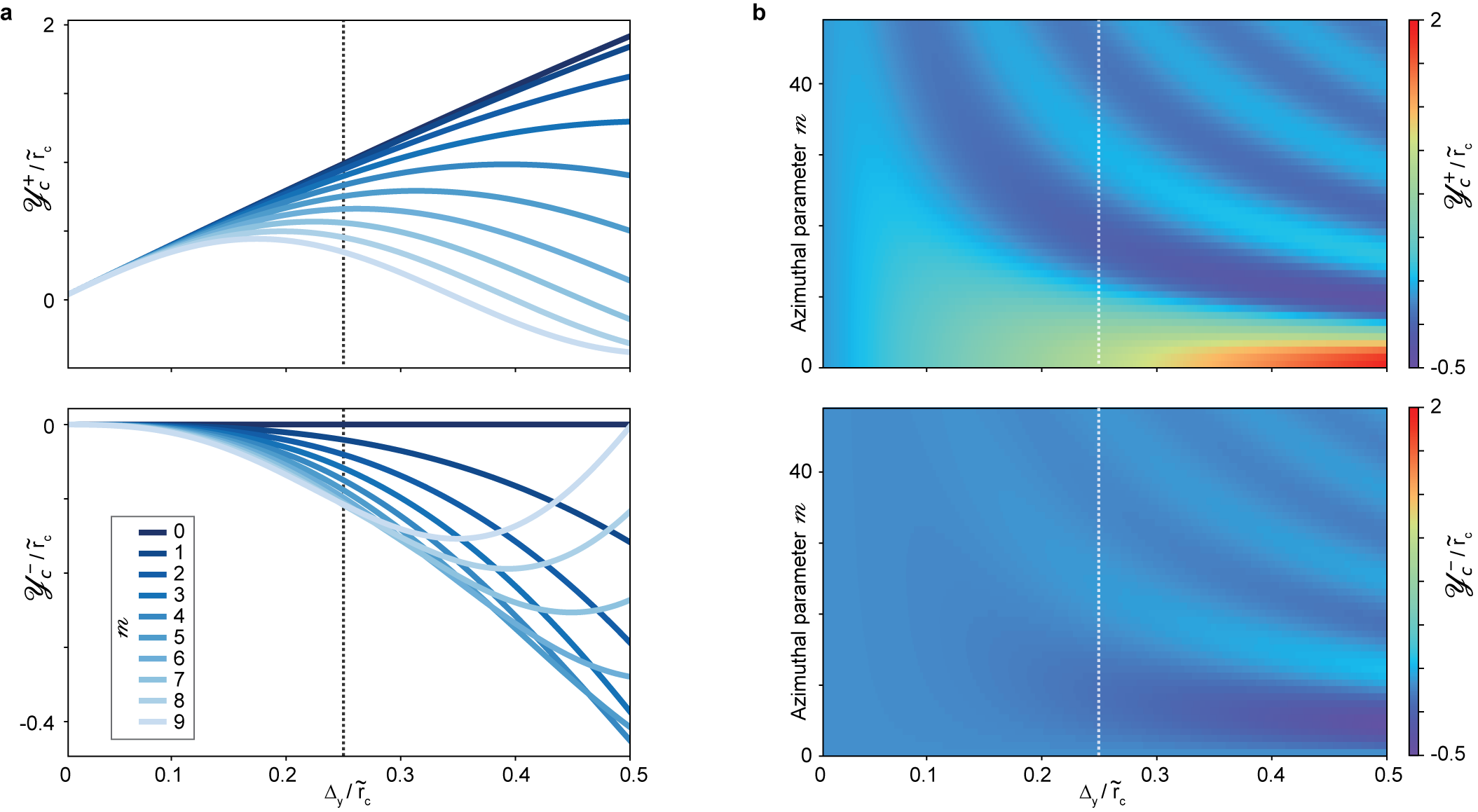}
\caption{\label{fig:Yc} \textbf{a}, Normalized amplitude variations of the induced flux with the lateral extension, when the coil is centered along the $\underline{y}$-direction following Eq.(\ref{eq:Yc}) in the limit of small lateral extension, for the ten first azimuthal parameter $\mathcal{m}$. \textbf{b,} Displayed as a color plot on an extended range of $\mathcal{m}$, quasiperiodic patterns at fixed $\Delta_y/\tilde{r}_c$ appear. The experimental setting ($\Delta_y = 0.5\,$mm for $\tilde{r}_c = 2\,$mm) is indicated by the vertical dashed lines.}
\end{figure*}

If the coil is centered regarding the $\underline{y}$-direction, $\Delta_y \equiv \Delta_y^+ = \Delta_y^-$ and 
\begin{equation}
\label{eq:Yc}
\mathcal{Y}_c^\pm = 2\Delta_y \left(\mathrm{sinc}\left[\frac{(\mathcal{m}+1)\Delta_y}{\tilde{r}_c}  \right] 
\pm  \mathrm{sinc} \left[\frac{(\mathcal{m}-1)\Delta_y}{\tilde{r}_c}  \right] \right)
\end{equation}
which is represented in Figure~\ref{fig:Yc}. While the signal amplitude globally decreases with $\mathcal{m}$, it exhibits oscillations which depends on the lateral extension $2\Delta_y$, the \textit{cylindrical} detection distance $\tilde{r}_c$ and $\mathcal{m}$, suggesting that $\tilde{r}_c$ can be adjusted to optimize the detection of a particular mode family.

The azimuthal dependence of a considered mode can be read in the phase of the measured magnetic flux, as given by Eq.(\ref{eq:flux}) as $\mathcal{m}\varphi_c$. These two lateral components $\mathcal{Y}^\pm$ control the contribution of the different components of the magnetic field to the flux envelope along $z$. 

\subsection{$\boldsymbol{z}$-direction}

\begin{figure*}
\includegraphics[height=\textheight,keepaspectratio]{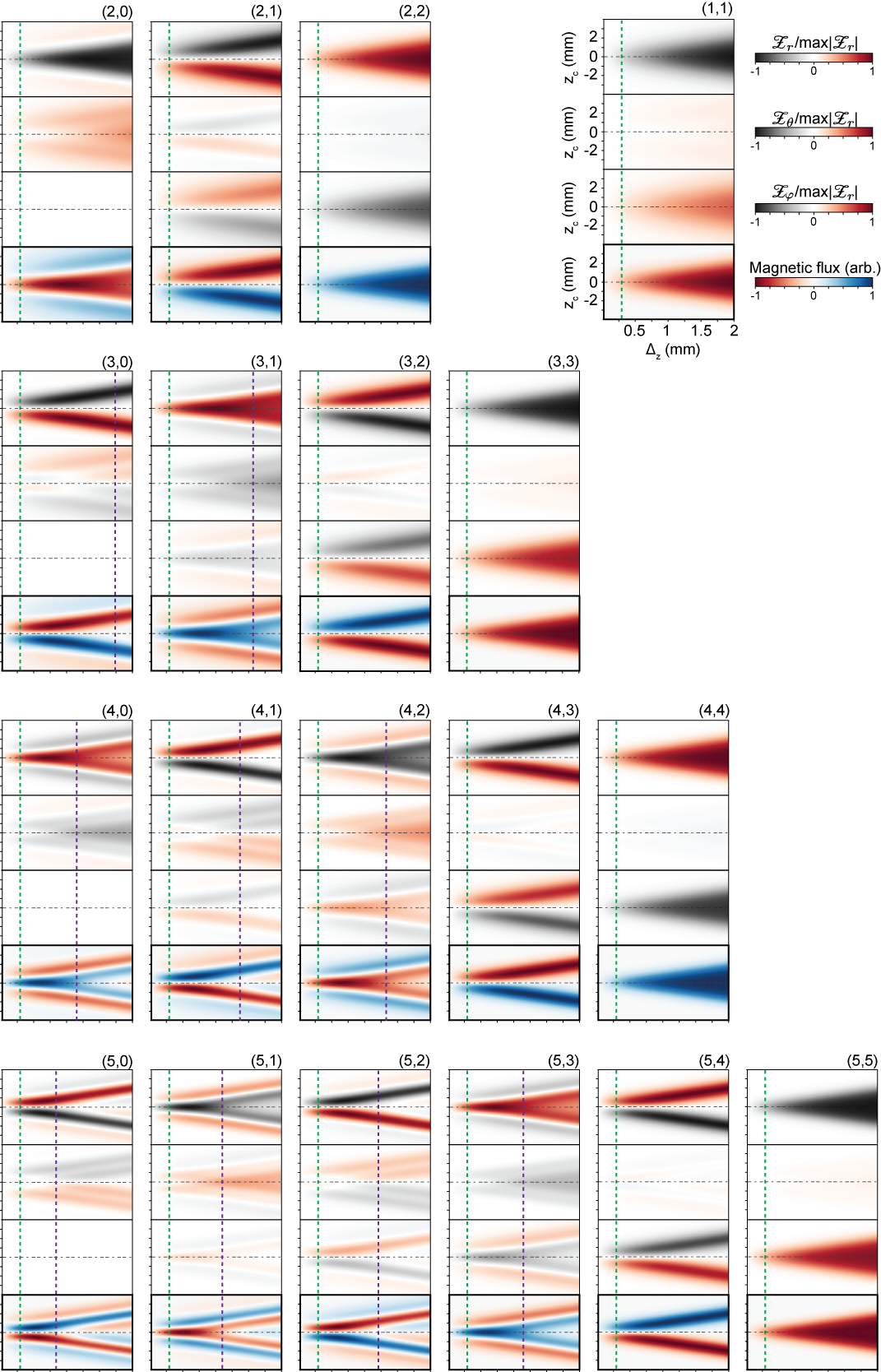}
\caption{\label{fig:Z} Decomposed magnetic flux envelopes along the $z$-axis as a function of the semi-extension of the coil height (in each panel, respectively $\mathcal{Z}_r$, $\mathcal{Z}_\theta$ and $\mathcal{Z}_\varphi$ from top to bottom) followed by the total magnetic flux described by Eq.(\ref{eq:flux}) for modes family with $\mathcal{n} < 6$ and $\mathcal{m} < 6$ (indicated on top right), calculated with the parameters $\Delta_y = 0.5$\,mm and $\tilde{r}_c = 2$\,mm. The experimental value $\Delta_z = 0.3$\,mm is indicated by green vertical dashed lines. The purple lines mark the apparition of artifacts due to an excessive height extension compared to the stray field polar features observed at this distance.}
\end{figure*}

Equations~(\ref{eq:Zr}), (\ref{eq:Ztheta}) and (\ref{eq:Zphi}) cannot be processed analytically. They can be simplified when working at a resonance such that $\alpha^\mathcal{m}_\mathcal{n} \gg \left(r/a\right)^{2\mathcal{n}+1}$, then 
\begin{equation}
 \Delta_1(r) \sim -(\mathcal{n}+1) \tilde{\Delta}_2(r) = -(\mathcal{n}+1) \frac{a^{2\mathcal{n}+1}}{r^{\mathcal{n}+2}} \, \alpha^\mathcal{m}_\mathcal{n} A^\mathcal{m}_\mathcal{n} \mathrm{.}
\end{equation}
The dependence of the flux $\phi^\mathcal{m}_\mathcal{n}$ with $1/\tilde{r}_c^{\mathcal{n}+2}$ along the equatorial plane partially conditions the choice of the detection distance. 
In Fig.~\ref{fig:Z} we represent, for the first spin-wave mode families with $\mathcal{n} < 6$ and $\mathcal{m} < 6$, the envelopes $\mathcal{Z}_r$, $\mathcal{Z}_\theta$ and $\mathcal{Z}_\varphi$ followed by the total magnetic flux $\phi^\mathcal{m}_\mathcal{n}$ along the altitude axis for different values of $\Delta_z$, the other parameters being fixed to their experimental values ($\Delta_y = 0.3$\, mm, $\tilde{r}_c = 2\,$mm). Once determined the azimuthal parameter $\mathcal{m}$, we access the polar parameter $\mathcal{n}-|\mathcal{m}|$ through the number of zeroes along the altitude $z$. 
The envelope along the $z$-axis is different for each mode, providing an additional identification method if the extent along $z$ experimentally accessible by the probe coil does not allow to go through all the envelope zeroes. 
The spin-wave mode polar structure with higher $\mathcal{n}-|\mathcal{m}|$ gets more and more complex, 
requiring the height dimension $2\Delta_z$ to be small enough for the flux not to be a mixture of its different features (marked informally by purple lines on Fig.~\ref{fig:Z}). This limit value accounts for $\Delta_z \sim 0.6$\,mm for $(7,0)$ and $(7,1)$ families. We design the coil such that $\Delta_z \sim 0.3\,$mm to comfortably work in the reliable region for at least $\mathcal{n}-\mathcal{m} \leq 7$. 
The complete magnetic fluxes computed in Fig.~1\textbf{c} and the insets of Fig.~4 in the main text result from the numerical evaluation of Eq.(\ref{eq:flux}) with the coil centered along $\underline{y}$ and the following parameters: $\Delta_y = 0.5\,$mm, $\Delta_z = 0.3\,$mm and $\tilde{r}_c = 2\,$mm.

\subsection{Non-ideality}
\subsubsection{Coil miscentered along the $\protect\underline{y}$-direction}
Evaluating the integral with $\Delta_y^- = \Delta_y^+ +\epsilon$ such that the total lateral extension is $2\Delta_y \equiv 2\Delta_y^+ +\epsilon$,
\begin{equation}
\mathcal{Y}_{\mathrm{mc}}^\pm = \mathcal{Y}_{\mathrm{c}}^\pm (\Delta_y^+) \pm 
\epsilon \left( e^{-i (\mathcal{m}-1) \frac{\Delta_y}{\tilde{r}_c}} \, \mathrm{sinc} \left[\frac{(\mathcal{m}-1) \epsilon}{2\tilde{r}_c} \right] \pm  
e^{-i (\mathcal{m}+1) \frac{\Delta_y}{\tilde{r}_c}} \, \mathrm{sinc} \left[\frac{(\mathcal{m}+1)\epsilon}{2\tilde{r}_c}\right] \right) 
\mathrm{.}
\end{equation}
At fixed altitude, this simply results in a constant extra phase, innocuous for the determination of $\mathcal{m}$. The phase variations between $\mathcal{Y}_\mathrm{mc}^+$ and $\mathcal{Y}_\mathrm{mc}^-$ weighted by the different envelopes $Z_r+Z_\theta$ and $Z_\varphi$ could induce a varying extra phase along the altitude axis. The effect could be neglected when $\mathcal{Z}_r +\mathcal{Z}_\theta \gg \mathcal{Z}_\varphi$. \color{black}

\subsubsection{Coil altitude-axis tilted compared to $\mathbf{H_\mathrm{DC}}$}
\begin{figure*}
\includegraphics[scale=1.2]{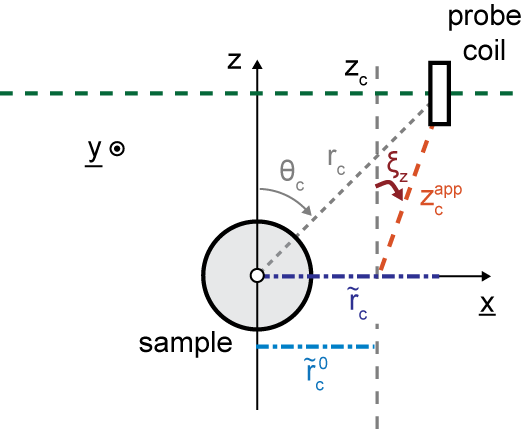}
\caption{\label{fig:repere_axe}Frame of the problem when the axis considered as the altitude degree of freedom of the probe coil is not parallel to the saturation axis defining the $z$-direction but tilted by an angle $\xi_z$ around the $\underline{y}$-axis.}
\end{figure*}
\paragraph*{}
A tilt of $\xi_z$ of the coil altitude-axis compared to the saturation axis (Figure~\ref{fig:repere_axe}) around $\underline{y}$ implies the coordinates $(\tilde{r}_c, z_c)$ injected in the calculations should be replaced by 
\begin{align*}
\tilde{r}_c &= \tilde{r}_{c}^0 + z_c^{\mathrm{app}} \, \sin \xi_z \mathrm{,}\\
z_c &= z_c^{\mathrm{app}} \, \cos \xi_z
\end{align*}
with $z_c^{\mathrm{app}}$ the apparent altitude along the coil $z_c$-axis and $\tilde{r}_{c}^0$ the \textit{cylindrical distance} at $z_c=0$.
The solid lines in Fig.~4 comes from the numerical evaluation of Eq.(\ref{eq:flux}) with the coil centered along $\underline{y}$ with $\Delta_y = 0.5\,$mm, $\Delta_z = 0.3\,$mm and $\tilde{r}_c = 2\,$mm and an axis tilt of $\xi_z = -6^\circ$. 
The axis tilt is determined by counterbalancing the predicted lobes amplitudes with respect to $z_c=0$ to match the detected modes. 

\section{Evolution of the eigenfrequencies with the static magnetic field $\mathbf{H}_\mathrm{DC}$}

\begin{figure*}
\includegraphics[scale=1]{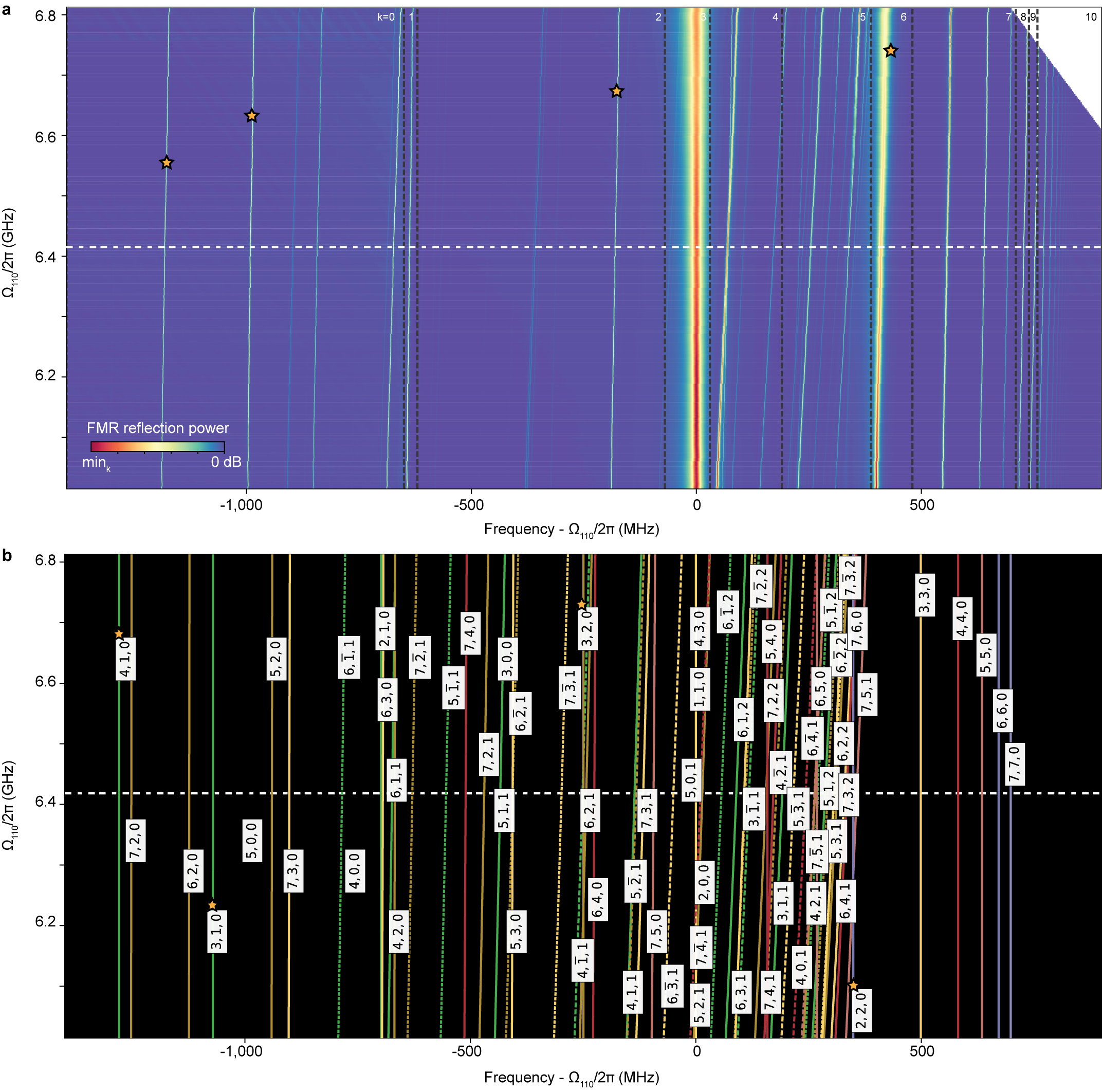} 
\caption{\label{fig:fI}\textbf{a,} Microwave reflection spectra from the pump coil as a function of the uniform mode precession frequency $\Omega_{110}/2\pi$ (vertical axis) --- tuned by varying the current flowing in the solenoid from $-0.8\,$A to $0.8\,$A. The uniform mode precession frequency also serves as a reference for the frequency horizontal axis. For readers' comfort, the colorscale has been adapted by frequency region (separated by black dashed lines and numbered by $k$ from 0 to 10) such that as many spin-wave modes as possible clearly appear ($\mathrm{min_k} = -0.48$, $-7.1$, $-2.5$, $-1.4$, $-2.1$, $-1.2$, $-2.3$, $-4.8$, $-0.91$, $-0.41$ and $-0.42\,$dB). \textbf{b,} Eigenfrequencies computed in the magnetostatic approximation from Eq.(\ref{resonance_eq}) for a magnetization saturation $M_s = 187\,$mT/$\mu_0$, $\gamma/2\pi=28\,$GHz/T up to $\mathcal{n} = 7$ and $\mathcal{m} = 7$. The orange stars mark the modes discussed in Fig.~4 in the main text. The white dashed line indicates the working point used in the main text with $\Omega_{110}/2\pi = 6.414\,$GHz.}
\end{figure*}

The evolution of the magnon eigenfrequencies with the static magnetic field ${H}_\mathrm{DC}$ can be predicted coarsely by solving the resonance condition in the magnetostatic approximation\cite{Walker1958,FletcherJAP1959}:
\begin{equation}
\label{resonance_eq}
\mathcal{n}+1 + \xi_0 \frac{P^{\mathcal{m}'}_\mathcal{n}(\xi_0)}{P^\mathcal{m}_\mathcal{n}(\xi_0)} \pm \mathcal{m}\nu = 0
\end{equation}
with
\begin{align*}
{\xi_0}^2 &= 1 + 1/\kappa,\\
\nu &= \frac{\Omega}{\Omega_H^2 - \Omega^2},\\
\kappa &= \frac{\Omega_H}{\Omega_H^2 - \Omega^2},
\end{align*}
the normalized internal field $\Omega_H=\frac{H_\mathrm{DC}- M_s/3}{M_s}$, the normalized magnon frequency $\Omega=\frac{\Omega_k}{\gamma \mu_0 M_s}$ and the gyromagnetic ratio $\gamma$.
We determine the effective saturation magnetization by comparing the frequency difference between (2,1,0) and (2,2,0), which are affected similarly by propagation effects~\cite{White1960}. In the magnetostatic approximation~\cite{FletcherJAP1959}, their frequencies $f_{210}$ and $f_{220}$ are such that
\begin{equation}
\mu_0 M_s = \frac{10 \pi}{\gamma} (f_{220}-f_{210})
\end{equation} 
which leads to an effective $M_s = 187\,$mT/$\mu_0$ in our experimental condition with $\Omega_{110}/2\pi = 6.414$\,GHz. 

\paragraph*{}
The solenoid wound around the magnetic circuit allows us to change the static magnetic field such that $\Omega_{110}/2\pi$ could be tuned from 6.01 to $6.81\,$GHz. The microwave reflection spectra are reported on Fig.~\ref{fig:fI}\textbf{a} above the computed eigenfrequencies expected in the magnetostatic approximation on Fig.~\ref{fig:fI}\textbf{b}.

\paragraph*{}
We compare the linespacing $2\pi \times \Delta f_{\mathcal{n}\mathcal{m}\mathcal{r}} = \Omega_{\mathcal{n}\mathcal{m}\mathcal{r}}-\Omega_{110}$ between the uniform precession mode (1,1,0) and others. In particular at the working point $\Omega_{110}/2\pi = 6.414$\,GHz, the modes discussed in Fig.~4 of the main text reading experimentally:
\begin{align*}
\Delta f_{220} &= +410\, \mathrm{MHz} \\
\Delta f_{320} &= -184\, \mathrm{MHz} \\
\Delta f_{310} &= -994\, \mathrm{MHz} \\
\Delta f_{410} &= -1185\, \mathrm{MHz} 
\end{align*}
are compared to the theoretical values predicted in the magnetostatic approximation, for $M_s = 187\,$mT/$\mu_0$:
\begin{align*}
\Delta f_{220} = &+349\, \mathrm{MHz} \\
\Delta f_{320} = &-249\, \mathrm{MHz} \\
\Delta f_{310} = &-1071	\, \mathrm{MHz} \\
\Delta f_{410} = &-1279\, \mathrm{MHz}
\end{align*}
resulting in the following relative discrepancies $14.9\%$, $-35.5\%$, $-7.7\%$ and $-7.9\%$.

\begin{figure*}
\includegraphics[scale=1]{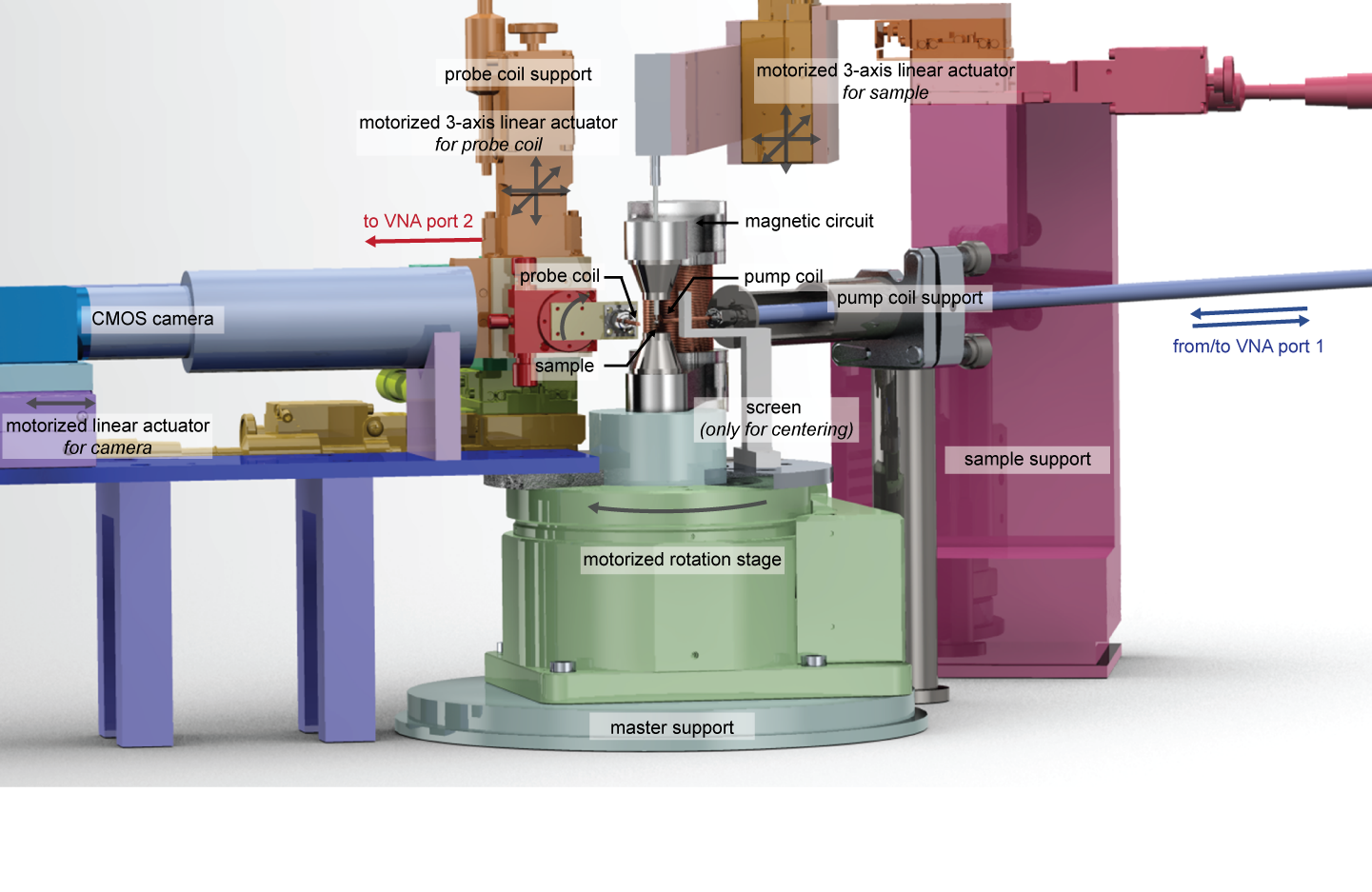}
\caption{\label{fig:setup}Setup of the core of the experiment.}
\end{figure*}
\section{Additional details on the setup}
\paragraph*{}
The core of the experimental setup is depicted on Fig.~\ref{fig:setup}. The motorized rotation stage and the magnetic circuit are screwed on the aluminum master support, clamped to the experiment table. The magnetic circuit is ended by conical concave elements in which can fit 12-mm diameter NdFeB permanent ring magnets. A solenoid is wound around the magnetic circuit to tune the static field ($\sim \pm 14\,$mT). 
The motorized rotation stage is an \textit{Optosigma} \textit{HST-120YAW} ($0.1^\circ$ position accuracy). The linear motors are composed by \textit{Optosigma} \textit{TAMM40-10C} (10$\,$mm travel range, $6\,\upmu$m position accuracy) and \textit{HPS60-20X-M5} ($20\,$mm travel range, $15\,\upmu$m position accuracy). 
The probe coil plane tilt is corrected with a small manual rotation stage \textit{Optosigma} \textit{KSP-256}.

The sample is attached to the end of a $\mathrm{Al}_2\mathrm{O}_3$ rod going through the upper permanent ring magnet, on a three-axis linear actuator. 
The centering of the sample with respect to the rotation stage axis is performed with the help of a CMOS camera, placed on the rotation stage to move jointly with the probe coil.
The CMOS camera is on a motorized translation stage --- to change dynamically the focus and evaluate distances --- fixed on the rotation stage. 
During the centering process, a white PVC screen with a carved target is fixed on the opposite side of the rotation stage. The sphere is centered along the rotation stage axis when its position is fixed on the target for any rotation stage angle. Designed to indicate the central position between the two magnets, it allows positioning the sample vertically in the static magnetic field. The screen is removed during the measurement to insure a maximum angular range.

\section{Data processing}
\begin{figure*}
\includegraphics[scale=1]{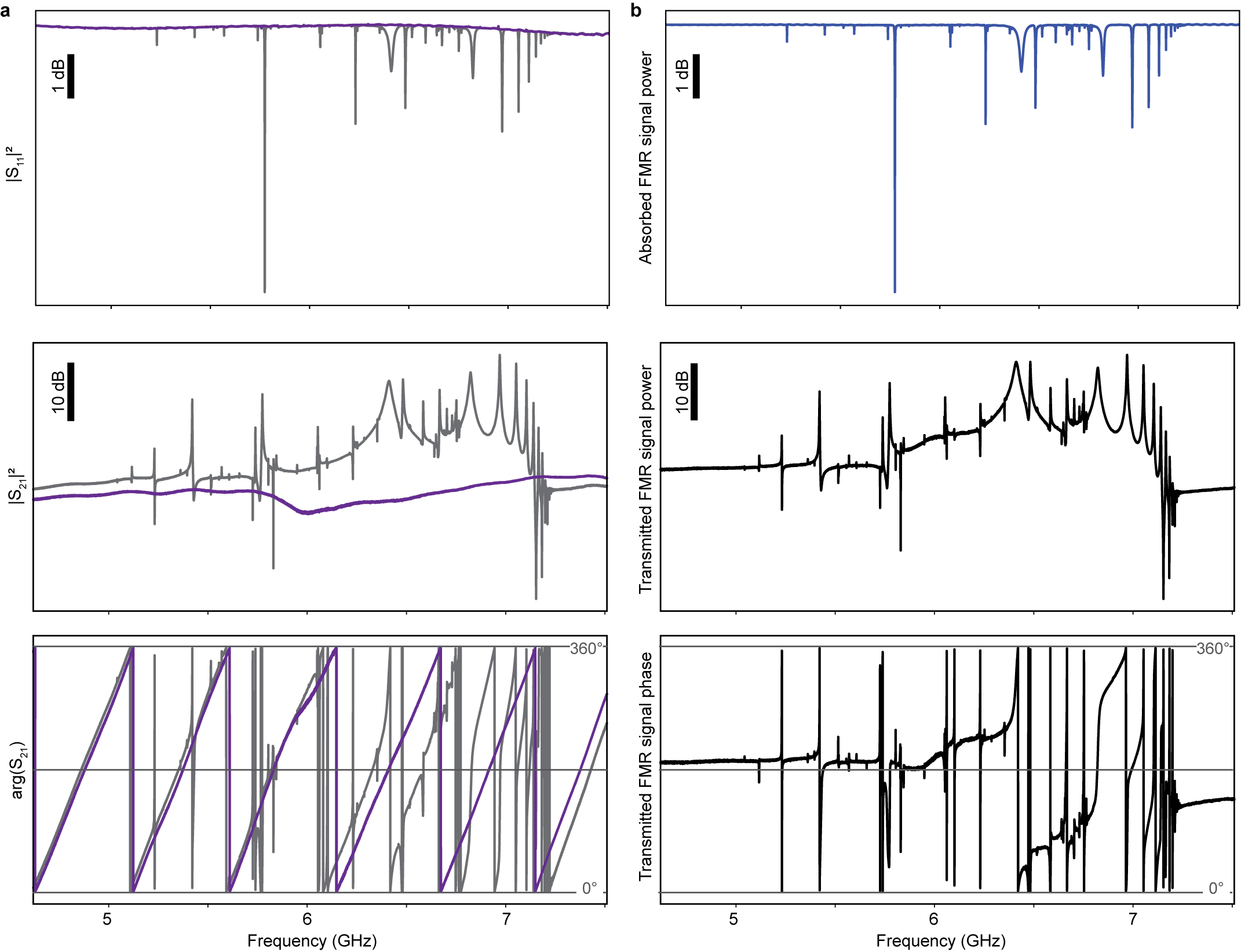}
\caption{\label{fig:calib}\textbf{a,} Typical raw spectra in reflection $S_{11}$ and in transmission $S_{21}$ with the sphere centered in the static magnetic field $H_\mathrm{DC}$ (grey lines) and vertically retracted (purple lines) as measured at a particular coil coordinate $(\varphi_c,z_c)$. \textbf{b,} Subsequent calibrated spectra obtained by normalizing the magnonic response with coils' local transfer function.}
\end{figure*}
\paragraph*{}
Defining solid phase references is an absolute necessity for this imaging method. At each coil position, we record the spectra ($S_{11}$ and $S_{21}$) in two configurations: with the sample in position centered in the static magnetic field and with the sample retracted vertically (Fig.~\ref{fig:calib}\textbf{a}). The latter gives access to the complex transfer functions of the pump coil and of the local pump-probe system which are used to normalize the magnonic spectra measured with the sample in position (Fig.~\ref{fig:calib}\textbf{b}).

\paragraph*{}
We gather hundreds of spectra from which we extract the spin-wave modes features, constituted by 67,242 points each across a span of 2.89$\,$GHz. These high-resolution spectral data require few hours of acquisition during which small temperature variations and the probe coil positions induce slight changes in the magnon frequencies and damping rates. We track and classify all the observable peaks in the dataset, before fitting them properly. Spectrum by spectrum, the reflection measurement (as in Fig.~2\textbf{a}) is used to get a precise guess of their eigenfrequencies by performing a Lorentzian fitting on a short sliding window of size comparable to the narrowest observed peak over the whole spectrum. 
The corresponding feature is considered as a plausible spin-wave mode if its fitted amplitude and linewidth fall within reasonable limits. This process leads to the pre-detection of tens of peaks. Comparing these peaks from spectrum to spectrum, the slightly deviating frequencies can be attributed to a particular mode by a nearest-neighbor approach. The process ends with a collection of eigenfrequencies over the whole dataset attributed to modes to be analyzed. 
\paragraph*{}
Information on the mode structures is contained in the transmission measurements. The phase of each mode is swamped in the background constituted by all the other modes. 
Great care is taken to adjust properly the response of the modes on each spectrum, progressing sequentially from the most dominant and correcting iteratively the fits to properly take into account the contribution of all the modes at a given frequency.

Once the amplitude of the dominant modes has been established, a position-dependent phenomenological background $S_0[\Omega] = (a \Omega^2 + b \Omega + c) e^{i (d+e\Omega)}$, likely due to remains of direct coupling and parasitic responses, could be piecewise-defined typically on $100\,$MHz, such that the solid red lines in Fig.~2\textbf{b} and Fig.~2\textbf{c} are respectively $|S_\phi[\Omega] + S_0[\Omega]|^2$ and $\mathrm{arg}(S_\phi[\Omega] + S_0[\Omega])$. This residual background with features much larger ($> 1$\,GHz) than the spin-wave modes ($< 10$\,MHz) allows to study modes appearing with small signal-to-noise ratio.

\section{Generalization to other shapes}
We generalize the technique by formally developing the expression of the magnetic flux induced by spin-wave modes in other ideal geometries. 
\subsection{Magnetic flux induced by a magnetostatic mode of a ferromagnetic cylinder}
\paragraph*{}
The outside magnetic potential induced by a magnetostatic mode $(\zeta,m)$ of a long axially-magnetized cylinder of radius $R$ is given by Joseph and Schl\"omann~\cite{Joseph1961}: 
\begin{equation}
\psi^\mathcal{m}_\mathcal{\zeta} (\tilde{r},\varphi, z) = A^\mathcal{m}_\mathcal{\zeta} K_\mathcal{m}(|\beta| {\tilde{r}}) e^{-i\beta z} e^{im\varphi}
\end{equation}
with  $\beta$ the propagation constant, $\zeta = |\beta|R$ and $K_\mathcal{m}$ the modified Bessel function of the second kind~\cite{Olver2010}.

The magnetic flux intercepted by a well-centered coil is 
\begin{equation}
\phi^\mathcal{m}_\mathcal{\xi} = \mu_0 \iint_S \, \boldsymbol{\nabla} \psi({\tilde{r}},\varphi,z) \cdot \mathbf{dS}_\mathcal{B} = \mu_0 \iint_S \, dS \left(I_{\tilde{r}} + I_\varphi \right)
\end{equation}
with 
\begin{align*}
I_{\tilde{r}} ({\tilde{r}},\underline{\varphi},z) &= e^{i \mathcal{m} \varphi_c} \left[e^{i (\mathcal{m}+1) \underline{\varphi}} +  e^{i (\mathcal{m}-1)\underline{\varphi}} \right] e^{-i \beta z} \Delta_{\tilde{r}}({\tilde{r}})    \mathrm{,} \\
I_\varphi(r,\theta, \underline{\varphi}) &= e^{i \mathcal{m} \varphi_c} \left[e^{i (\mathcal{m}+1) \underline{\varphi}} -  e^{i (\mathcal{m}-1)\underline{\varphi}} \right]
e^{-i \beta z}\Delta_\varphi({\tilde{r}})   
\end{align*}

and
\begin{align*}
\Delta_{\tilde{r}}({\tilde{r}}) =&  -\frac{|\beta|}{4}\left[K_\mathcal{m-1}(|\beta| {\tilde{r}}) + K_\mathcal{m+1}(|\beta| {\tilde{r}})\right] \\
\Delta_\varphi({\tilde{r}}) =&  -\frac{m}{2{\tilde{r}}} K_\mathcal{m}(|\beta| {\tilde{r}}) 
\end{align*}

Considering $\mathcal{Y}^\pm$ previously presented, it can all be compacted as 
\begin{equation}
\phi^\mathcal{m}_\mathcal{\xi} = \mu_0 e^{i \mathcal{m} \varphi_c} e^{-i \beta z_c} \,\, \big( 2\Delta_z \mathrm{sinc}(\beta \Delta_z) \, \left[\mathcal{Y}^+ \Delta_{\tilde{r}}(\tilde{r}_c) + \mathcal{Y}^-   \Delta_\varphi(\tilde{r}_c) \right] \big) \mathrm{.}
\end{equation}
The presented imaging technique can be directly applied to cylinders by reading the azimuthal and the altitude dependences in the phase evolution along the cylindrical orbital of radius $\tilde{r}_c$.

\subsection{Magnetic flux induced by a magnetostatic mode of a spheroid}
\paragraph*{}
Following the description given by Walker~\cite{Walker1957} in oblate spheroidal coordinates $(\xi, \eta,\varphi)$, the magnetic potential outside an oblate spheroid of transverse semi-axis $a$ and longitudinal semi-axis $b$ reads:  
\begin{equation}
\psi^\mathcal{m}_\mathcal{n} (\xi,\eta, \varphi) =  Q^m_n(i\xi) \, P^m_n(\eta) \, e^{i m\varphi}
\end{equation}
with $Q^m_n$ the Ferrers function of the second kind~\cite{Olver2010} and defining $\iota^2 \equiv a^2-b^2$, 
\begin{align}
x &= \iota \, (1+\xi^2)^{1/2} \,  (1-\eta^2)^{1/2} \,  \cos \varphi \mathrm{,}\\
y &= \iota \, (1+\xi^2)^{1/2} \,  (1-\eta^2)^{1/2} \,  \sin \varphi \mathrm{,}\\
z &= \iota \, \xi \, \eta \mathrm{.}
\end{align}

\begin{figure*}
\includegraphics[scale=0.9]{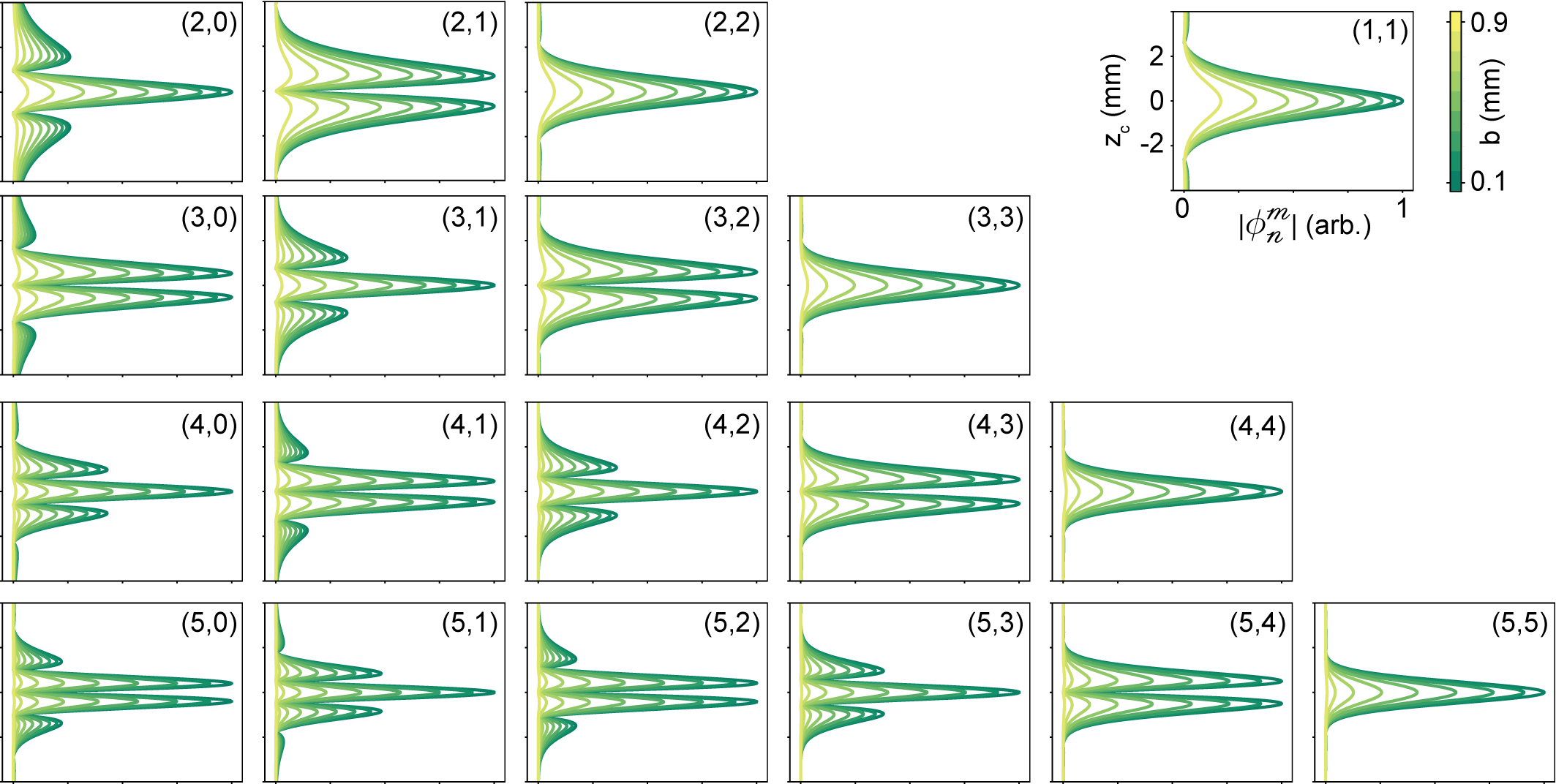}
\caption{\label{fig:Zell}
Norm of the induced magnetic flux by a magnetized spheroid described by Eq.~(\ref{eq:phiell}) computed for modes family with $\mathcal{n} < 6$ and $\mathcal{m} < 6$, in the experimental conditions ($\Delta_y = 0.5$\,mm,  $\Delta_z = 0.3$\,mm, $\tilde{r}_c = 2$\,mm, $a = 1$\,mm) for values of the longitudinal semi-axis $b$ varying from 0.1\,mm (dark green) to 0.9\,mm (yellow). For each mode ($\mathcal{n},\mathcal{m}$), the data are normalized by the maximum in $b = 0.1$\,mm. Approaching the sphere limit when $b \sim a$, the pattern is the same as observed in Sec.~\ref{section:theo}.
}
\end{figure*} 

These oblate coordinates can be expressed in cylindrical coordinates:
\begin{align}
\xi^2({\tilde{r}},z) &= \, \frac{1}{2\iota^2} \,\,\left[\Lambda({\tilde{r}},z) + \Upsilon({\tilde{r}},z)\right] \mathrm{,} \\
\eta({\tilde{r}},z) &= \sqrt{2} \, z \left[\Lambda({\tilde{r}},z) + \Upsilon({\tilde{r}},z)\right]^{-\frac{1}{2}} \mathrm{.}
\end{align}

with 
\begin{align}
\Lambda({\tilde{r}},z) &= \tilde{r}^2 + z^2 - \iota^2  \mathrm{,}\\
\Upsilon({\tilde{r}},z) &= \left[ 4 \iota^2 z^2 + \Lambda^2 \right]^{\frac{1}{2}} \mathrm{.}
\end{align}

With a well-centered coil,
\begin{equation}
\phi^\mathcal{m}_\mathcal{n} = \mu_0 \iint_S \, \boldsymbol{\nabla} \psi({\tilde{r}},\varphi, z) \cdot \mathbf{dS}_\mathcal{B} = \mu_0 \iint_S \,dS \,(I_{\tilde{r}} + I_\varphi)
\end{equation}
with 
\begin{align}
I_{\tilde{r}} =& \,  e^{i m\varphi_c} \,  \left[e^{i (\mathcal{m}+1) \underline{\varphi}} +  e^{i (\mathcal{m}-1)\underline{\varphi}} \right]  \,\,\,\, \frac{1}{2} \partial_{\tilde{r}}[Q^m_n(i\xi) \, P^m_n(\eta)] \mathrm{,}\\
I_\varphi =& \,  e^{i m\varphi_c} \, \left[e^{i (\mathcal{m}+1) \underline{\varphi}} -  e^{i (\mathcal{m}-1)\underline{\varphi}} \right]  \,\,\,\, \frac{-m}{2{\tilde{r}}} \, Q^m_n(i\xi) \, P^m_n(\eta) 
\end{align}
so that 
\begin{equation}
\label{eq:phiell}
\phi^\mathcal{m}_\mathcal{n} = \mu_0 e^{i m \varphi_c} \left[\mathcal{Y}^+ \mathcal{Z}_{\tilde{r}}(z_c) - \mathcal{Y}^- \mathcal{Z}_\varphi(z_c) \right]\mathrm{.}
\end{equation}
The dependence along the azimuth is the same as seen previously. 
Along the altitude, 
\begin{align}
\mathcal{Z}_{\tilde{r}}(z_c) &\equiv  \frac{1}{2} \int_{z_c-\Delta_z^-}^{z_c+\Delta_z^+} dz \,  \partial_{\tilde{r}}[Q^m_n(i\xi[\tilde{r}_c,z]) \, P^m_n(\eta[\tilde{r}_c,z] )] \mathrm{,} \\
\mathcal{Z}_\varphi(z_c)    &\equiv  \frac{m}{2\tilde{r}_c} \int_{z_c-\Delta_z^-}^{z_c+\Delta_z^+} dz \,    \, Q^m_n(i\xi[\tilde{r}_c,z]) \, P^m_n(\eta[\tilde{r}_c,z]) 
\end{align}

with the derivatives along ${\tilde{r}}$ reading
\begin{align}
\partial_{\tilde{r}} \xi({\tilde{r}},z) &=  \,\,\,\frac{1}{\sqrt{2}{\iota}}\,\,\, {\tilde{r}} \,\, \frac{[\Lambda({\tilde{r}},z) + \Upsilon({\tilde{r}},z)]^{\frac{1}{2}}}{\Upsilon({\tilde{r}},z)}\mathrm{,}\\
\partial_{\tilde{r}} \eta({\tilde{r}},z) &= -\sqrt{2} \,\, z{\tilde{r}} \,\, \frac{[\Lambda({\tilde{r}},z) + \Upsilon({\tilde{r}},z) ]^{-\frac{1}{2}}}{\Upsilon({\tilde{r}},z)}
\end{align}
and 
\begin{align}
\partial_{\tilde{r}} [Q^m_n(i\xi) \, P^m_n(\eta)] =& \frac{\partial_{\tilde{r}} \eta}{\eta^2-1} Q^m_n(i\xi) \left[-(n+1) P^m_n(\eta) \, \eta + (n-m+1) P^m_{n+1}(\eta)\right] \\
-& \frac{i\partial_{\tilde{r}} \xi}{\xi^2+1} P^m_n(\eta) \, \left[-(n+1) Q^m_n(i\xi) \, i\xi + (n-m+1) Q^m_{n+1}(i\xi)\right] \mathrm{.}
\end{align}

These expressions can be numerically computed to provide analog patterns to Fig.~\ref{fig:Z} and be similarly used for identification. For illustration, in Fig.~\ref{fig:Zell} we plot the norm of the flux that would be intercepted for the first modes in our experimental conditions for various values of ellipticity. These developments suggest the imaging method can be directly applied to any spheroid.